\documentclass[11pt]{article}

\usepackage[dvips]{graphicx}
\usepackage{float}
\usepackage{epsfig}
\usepackage{ulem}
\usepackage{latexsym,amsmath,amsfonts,amssymb}
\usepackage[latin1]{inputenc}

\usepackage{rotating}
\usepackage[american]{babel}
\usepackage[dvips]{graphicx}
\usepackage{bbm}
\usepackage{color}
\usepackage{slashed}
\usepackage[unicode]{hyperref}
\usepackage{lscape}
\usepackage{bigints}
\usepackage{enumerate}
\usepackage[shortlabels]{enumitem}

\usepackage{tikz}
\usetikzlibrary{decorations.pathreplacing}
\usetikzlibrary{shapes}
\usepackage{graphicx}
\usepackage{epsfig}
\usepackage{rotating}
\usepackage{amssymb}
\usepackage{subfigure}
\usepackage{dsfont}
\usepackage{psfrag}
\usepackage{amsmath,euscript,array,mathrsfs,amsfonts}
\usepackage{slashed}
\usepackage{array}
\usepackage{youngtab}
\usepackage{color}
\usepackage{bbold}

\usepackage{hyperref}
\hypersetup{
colorlinks = true,
linkcolor = red,
linktocpage = true,
citecolor = blue
}

\usepackage{tikz}
\usetikzlibrary{calc}
 \usetikzlibrary{decorations.text}
 \usetikzlibrary{shapes}

 \usetikzlibrary{decorations.pathmorphing}
\usetikzlibrary{decorations.pathreplacing}
\usetikzlibrary{arrows.meta}
\tikzset{
  >={To[length=5pt]}
  }
\usetikzlibrary{shapes, shapes.geometric, shapes.symbols, shapes.arrows, shapes.multipart, shapes.callouts, shapes.misc}
\tikzset{snake it/.style={decorate, decoration=snake}}
\tikzset{7brane/.style={circle, draw=black, fill=black,ultra thick,inner sep=1.5 pt, minimum size=1 pt,}, c/.default={4pt}}
\tikzset{cross/.style={cross out, draw=black,thick, minimum size=2*(#1-\pgflinewidth), inner sep=0pt, outer sep=0pt}, cross/.default={5pt}}
\tikzset{big7brane/.style={circle, draw=black, fill=black,ultra thick,inner sep=2.5 pt, minimum size=1 pt,}, c/.default={4pt}}
\tikzset{u/.style={circle, draw=black, fill=white,inner sep=2 pt, minimum size=2 pt,},f/.style={square, draw=black, fill=white,ultra thick,inner sep=4 pt, minimum size=2 pt,}}
\tikzset{so/.style={circle, draw=black, fill=red,inner sep=2 pt, minimum size=2 pt,},f/.style={square, draw=black, fill=white,ultra thick,inner sep=4 pt, minimum size=2 pt,}}
\tikzset{sp/.style={circle, draw=black, fill=blue,inner sep=2 pt, minimum size=2 pt,},f/.style={square, draw=black, fill=white,ultra thick,inner sep=4 pt, minimum size=2 pt,}}
\tikzset{uf/.style={rectangle, draw=black, fill=white,inner sep=3 pt, minimum size=4 pt,}}
\tikzset{spf/.style={rectangle, draw=black, fill=blue, thick,inner sep=3 pt, minimum size=4 pt, circle, draw=black, fill=blue,thick,inner sep=2 pt, minimum size=2 pt,},f/.style={square, draw=black, fill=white,ultra thick,inner sep=4 pt, minimum size=2 pt,}}
\tikzset{sof/.style={rectangle, draw=black, fill=red, thick,inner sep=3 pt, minimum size=4 pt,}}
\usetikzlibrary{positioning}
\usetikzlibrary{arrows}
\usetikzlibrary{decorations.pathreplacing}
\usetikzlibrary{shapes}

\makeatletter\def\l@subsubsection#1#2{}%
\makeatother

\renewcommand\theequation{\arabic{section}.\arabic{equation}} 

\usepackage{subfigure}

\def\CC{\ensuremath{\mathds C}}
\def\RR{\ensuremath{\mathds R}}
\def\ZZ{\ensuremath{\mathds Z}}

\DeclareMathOperator{\vol}{vol}
\DeclareMathOperator{\Vol}{Vol}
\DeclareMathOperator{\sech}{sech}
\DeclareMathOperator{\tr}{tr}

\DeclareMathOperator{\csch}{csch}

\newcommand{\be}{\begin{equation}}
\newcommand{\ee}{\end{equation}}
\newcommand{\ba}{\begin{array}}
\newcommand{\ea}{\end{array}}

\newcommand{\htheta}{\hat{\theta}}
\newcommand{\hphi}{\hat{\phi}}
\newcommand{\hr}{\hat{r}}
\newcommand{\hm}{\hat{m}}

\def\im{Invent. Math.}

\def\hat{\widehat}
\def\a{\alpha}
\def\b{\beta}
\def\c{\gamma}
\def\d{\delta}
\def\f{\phi}               
\def\vf{\varphi}  
\def\tvf{\tilde{\varphi}}
\def\vp{\varphi}
\def\g{\gamma}
\def\h{\eta}
\def\j{\psi}
\def\k{\kappa}                    
\def\l{\lambda}
\def\m{\mu}
\def\n{\nu}
\def\o{\omega}  \def\w{\omega}

\def\q{\theta}  \def\th{\theta}                  
\def\r{\rho}                                     
\def\s{\sigma}                                   
\def\t{\tau}
\def\u{\upsilon}
\def\x{\xi}
\def\z{\zeta}
\def\pt{\tilde{\varphi}}
\def\tt{\tilde{\theta}}
\def\lab{\label}
\def\6{\partial}
\def\wg{\wedge}
\def\bpsi{\bar{\psi}}
\def\bt{\bar{\theta}}
\def\bvf{\bar{\varphi}}

\newcommand{\beq}{\begin{equation}}
\newcommand{\eeq}{\end{equation}}
\newcommand{\bea}{\begin{eqnarray}}
\newcommand{\eea}{\end{eqnarray}}

\newcommand{\beqs}{\begin{eqnarray}}
\newcommand{\eeqs}{\end{eqnarray}}
\newcommand{\bal}{\begin{aligned}}
\newcommand{\eal}{\end{aligned}}
\makeatletter
\newcommand\setItemnumber[1]{\setcounter{enum\romannumeral\@enumdepth}{\numexpr#1-1\relax}}
\makeatother
\begin{document}
\baselineskip=15.5pt
\pagestyle{plain}
\setcounter{page}{1}

\def\del{{\partial}}
\def\vev#1{\left\langle #1 \right\rangle}
\def\cn{{\cal N}}
\def\co{{\cal O}}


\def\IC{{\mathbb C}}
\def\IR{{\mathbb R}}
\def\IZ{{\mathbb Z}}
\def\RP{{\bf RP}}
\def\CP{{\bf CP}}
\def\Poincaré{{Poincar\'e }}
\def\tr{{\rm tr}}
\def\tp{{\tilde \Phi}}

\newcommand{\cotg}{\text{cotg}}
\newcommand{\M}{\mathcal{M}}

\def\TL{\hfil$\displaystyle{##}$}
\def\TR{$\displaystyle{{}##}$\hfil}
\def\TC{\hfil$\displaystyle{##}$\hfil}
\def\TT{\hbox{##}}
\def\HLINE{\noalign{\vskip1\jot}\hline\noalign{\vskip1\jot}}
\def\seqalign#1#2{\vcenter{\openup1\jot
   \halign{\strut #1\cr #2 \cr}}}
\def\lbldef#1#2{\expandafter\gdef\csname #1\endcsname {#2}}
\def\eqn#1#2{\lbldef{#1}{(\ref{#1})}%
\begin{equation} #2 \label{#1} \end{equation}}
\def\eqalign#1{\vcenter{\openup1\jot
     \halign{\strut\span\TL & \span\TR\cr #1 \cr
    }}}

\def\eno#1{(\ref{#1})}
\def\href#1#2{#2}
\def\half{\frac{1}{2}}



\def\ads{{\it AdS}}
\def\adsp{{\it AdS}$_{p+2}$}
\def\cft{{\it CFT}}

\newcommand{\ber}{\begin{eqnarray}}
\newcommand{\eer}{\end{eqnarray}}

\newcommand{\beqar}{\begin{eqnarray}}
\newcommand{\cO}{{\cal O}}
\newcommand{\cT}{{\cal T}}
\newcommand{\cR}{{\cal R}}
\newcommand{\eeqar}{\end{eqnarray}}
\newcommand{\tht}{\thteta}
\newcommand{\lm}{\lambda}\newcommand{\Lm}{\Lambda}


\newcommand{\nonu}{\nonumber}
\newcommand{\oh}{\displaystyle{\frac{1}{2}}}
\newcommand{\dsl}
   {\kern.06em\hbox{\raise.15ex\hbox{$/$}\kern-.56em\hbox{$\partial$}}}
\newcommand{\as}{\not\!\! A}
\newcommand{\ps}{\not\! p}
\newcommand{\ks}{\not\! k}
\newcommand{\D}{{\cal{D}}}
\newcommand{\dv}{d^2x}
\newcommand{\Z}{{\cal Z}}
\newcommand{\N}{{\cal N}}
\newcommand{\Dsl}{\not\!\! D}
\newcommand{\Bsl}{\not\!\! B}
\newcommand{\Psl}{\not\!\! P}

\newcommand{\eeqarr}{\end{eqnarray}}


\def\del{{\delta^{\hbox{\sevenrm B}}}} \def\ex{{\hbox{\rm e}}}
\def\azb{A_{\bar z}} \def\az{A_z} \def\bzb{B_{\bar z}} \def\bz{B_z}
\def\czb{C_{\bar z}} \def\cz{C_z} \def\dzb{D_{\bar z}} \def\dz{D_z}
\def\im{{\hbox{\rm Im}}} \def\mod{{\hbox{\rm mod}}} \def\tr{{\hbox{\rm Tr}}}
\def\ch{{\hbox{\rm ch}}} \def\imp{{\hbox{\sevenrm Im}}}
\def\trp{{\hbox{\sevenrm Tr}}} \def\vol{{\hbox{\rm Vol}}}
\def\rl{\Lambda_{\hbox{\sevenrm R}}} \def\wl{\Lambda_{\hbox{\sevenrm W}}}
\def\fc{{\cal F}_{k+\cox}} \def\vev{vacuum expectation value}
\def\nodiv{\mid{\hbox{\hskip-7.8pt/}}}
\def\ie{{\em i.e.}}
\def\ie{\hbox{\it i.e.}}

\def\CC{{\mathchoice
{\rm C\mkern-8mu\vrule height1.45ex depth-.05ex
width.05em\mkern9mu\kern-.05em}
{\rm C\mkern-8mu\vrule height1.45ex depth-.05ex
width.05em\mkern9mu\kern-.05em}
{\rm C\mkern-8mu\vrule height1ex depth-.07ex
width.035em\mkern9mu\kern-.035em}
{\rm C\mkern-8mu\vrule height.65ex depth-.1ex
width.025em\mkern8mu\kern-.025em}}}

\def\RR{{\rm I\kern-1.6pt {\rm R}}}
\def\NN{{\rm I\!N}}
\def\ZZ{{\rm Z}\kern-3.8pt {\rm Z} \kern2pt}
\def\IB{\relax{\rm I\kern-.18em B}}
\def\ID{\relax{\rm I\kern-.18em D}}
\def\II{\relax{\rm I\kern-.18em I}}
\def\IP{\relax{\rm I\kern-.18em P}}
\newcommand{\CS}{{\scriptstyle {\rm CS}}}
\newcommand{\CSs}{{\scriptscriptstyle {\rm CS}}}
\newcommand{\rc}{\nonumber\\}
\newcommand{\bear}{\begin{eqnarray}}
\newcommand{\eear}{\end{eqnarray}}

\newcommand{\LL}{{\cal L}}

\def\mani{{\cal M}}
\def\calo{{\cal O}}
\def\calb{{\cal B}}
\def\calw{{\cal W}}
\def\calz{{\cal Z}}
\def\cald{{\cal D}}
\def\calc{{\cal C}}

\def\to{\rightarrow}
\def\ele{{\hbox{\sevenrm L}}}
\def\ere{{\hbox{\sevenrm R}}}
\def\zb{{\bar z}}
\def\wb{{\bar w}}
\def\nodiv{\mid{\hbox{\hskip-7.8pt/}}}
\def\menos{\hbox{\hskip-2.9pt}}
\def\dr{\dot R_}
\def\drr{\dot r_}
\def\ds{\dot s_}
\def\da{\dot A_}
\def\dga{\dot \gamma_}
\def\ga{\gamma_}
\def\dal{\dot\alpha_}
\def\al{\alpha_}
\def\cl{{closed}}
\def\cls{{closing}}
\def\vev{vacuum expectation value}
\def\tr{{\rm Tr}}
\def\to{\rightarrow}
\def\too{\longrightarrow}


\def\a{\alpha}
\def\b{\beta}
\def\c{\gamma}
\def\d{\delta}
\def\e{\epsilon}           
\def\F{\Phi}
\def\f{\phi}               
\def\vf{\varphi}  \def\tvf{\tilde{\varphi}}
\def\vp{\varphi}
\def\g{\gamma}
\def\h{\eta}
\def\j{\psi}
\def\k{\kappa}                    
\def\l{\lambda}
\def\m{\mu}
\def\n{\nu}
\def\o{\omega}  \def\w{\omega}
\def\q{\theta}  \def\th{\theta}                  
\def\r{\rho}                                     
\def\s{\sigma}                                   
\def\t{\tau}
\def\u{\upsilon}
\def\x{\xi}
\def\X{\Xi}
\def\z{\zeta}
\def\pt{\tilde{\varphi}}
\def\tt{\tilde{\theta}}
\def\lab{\label}
\def\6{\partial}
\def\wg{\wedge}
\def\atanh{{\rm arctanh}}
\def\bpsi{\bar{\psi}}
\def\bt{\bar{\theta}}
\def\bvf{\bar{\varphi}}

%



\newfont{\namefont}{cmr10}
\newfont{\addfont}{cmti7 scaled 1440}
\newfont{\boldmathfont}{cmbx10}
\newfont{\headfontb}{cmbx10 scaled 1728}





\newcommand{\re}{\,\mathbb{R}\mbox{e}\,}
\newcommand{\hyph}[1]{$#1$\nobreakdash-\hspace{0pt}}
\providecommand{\abs}[1]{\lvert#1\rvert}
\newcommand{\Nugual}[1]{$\mathcal{N}= #1 $}
\newcommand{\sub}[2]{#1_\text{#2}}
\newcommand{\partfrac}[2]{\frac{\partial #1}{\partial #2}}
\newcommand{\bsp}[1]{\begin{equation} \begin{split} #1 \end{split} \end{equation}}
\newcommand{\calF}{\mathcal{F}}
\newcommand{\calO}{\mathcal{O}}
\newcommand{\calM}{\mathcal{M}}
\newcommand{\calV}{\mathcal{V}}
\newcommand{\bbZ}{\mathbb{Z}}
\newcommand{\bbC}{\mathbb{C}}
\newcommand{\cK}{{\cal K}}

\newcommand{\Thq}{\Theta\left(\r-\r_q\right)}
\newcommand{\Dq}{\d\left(\r-\r_q\right)}
\newcommand{\kten}{\kappa^2_{\left(10\right)}}
\newcommand{\pbi}[1]{\imath^*\left(#1\right)}
\newcommand{\ho}{\hat{\omega}}
\newcommand{\tth}{\tilde{\th}}
\newcommand{\tf}{\tilde{\f}}
\newcommand{\tj}{\tilde{\j}}
\newcommand{\tw}{\tilde{\omega}}
\newcommand{\tz}{\tilde{z}}
\newcommand{\prj}[2]{(\partial_r{#1})(\partial_{\j}{#2})-(\partial_r{#2})(\partial_{\j}{#1})}
\def\atanh{{\rm arctanh}}
\def\sech{{\rm sech}}
\def\csch{{\rm csch}}
\allowdisplaybreaks[1]

\def\red{\textcolor[rgb]{0.98,0.00,0.00}}

\newcommand{\Dan}[1] {{\textcolor{blue}{#1}}}

\numberwithin{equation}{section}

\newcommand{\Tr}{\mbox{Tr}}    


%

\setcounter{footnote}{0}
\renewcommand{\theequation}{{\rm\thesection.\arabic{equation}}}

\begin{titlepage}

\begin{center}

\vskip .5in 
\noindent

{\Large \bf{ Compactification of 6d ${\cal N}=(1,0)$ quivers, 4d SCFTs and their holographic dual Massive IIA backgrounds} }
\bigskip\medskip

Paul Merrikin \footnote{paulmerrikin@hotmail.co.uk, p.r.g.merrikin.2043506@swansea.ac.uk}, Carlos Nunez\footnote{c.nunez@swansea.ac.uk} and Ricardo Stuardo\footnote{ricardostuardotroncoso@gmail.com}\\

\bigskip\medskip
{\small 
Department of Physics, Swansea University, Swansea SA2 8PP, United Kingdom}

\vskip .5cm 
\vskip .9cm 
     	{\bf Abstract }\vskip .1in
\end{center}

\noindent
In this paper we study an infinite family of Massive Type IIA backgrounds that holographically describe the twisted compactification of ${\cal N}=(1,0)$ six-dimensional SCFTs to four dimensions. The analysis of the branes involved motivates an {\it heuristic} proposal for a four dimensional linear quiver QFT, that deconstructs the theory in six dimensions. For the case in which the system reaches a strongly coupled fixed point, we calculate some observables that we compare with holographic results. Two quantities measuring the number of degrees of freedom for the flow across dimensions are studied.
 
 \noindent
\vskip .5cm
\vskip .5cm
\vfill
\eject

\end{titlepage}

\setcounter{footnote}{0}

\small{
\tableofcontents}

\normalsize

\newpage
\renewcommand{\theequation}{{\rm\thesection.\arabic{equation}}}
\section{Introduction}

Maldacena's AdS/CFT conjecture \cite{Maldacena:1997re} motivates the study of both gravity and field theory  topics. In particular, the study of supersymmetric and conformal field theories in diverse dimensions. 
In the past few years we witnessed the definition of new, characteristically non-Lagrangian CFTs, by the existence of a trustable background of Type II or M-theory, containing an AdS-factor.

In fact, this procedure has been applied to the possible space-time dimensions for which super conformal field theories exist ($d+1=1,....,6$). With eight Poincare supercharges, there exists  a classification and an algorithmic way of associating a particular SCFT$_{d+1}$ with a Type II background containing an AdS$_{d+2}$ factor. At present there seems to be exceptions to this statement for the cases of supergravity solutions containing AdS$_3$ and AdS$_2$ spaces. See \cite{oned}-\cite{Nunez:2018ags}, for references working details of the cases $(d+1)=1,2,3,4,5,6$. A comprehensive summary of  various aspects of SCFTs in diverse dimensions can be found in \cite{Argyres:2022mnu}.

A reasonable extension is the study of RG-flows away from these SCFTs$_{d+1}$. These flows can be between two conformal points or between a CFT and a gapped theory. Less conventional are the flows across dimensions, between a SCFT$_{D+1}$ and a SCFT$_{d+1}$ (there is also  with the possibility of ending in gapped systems). There are numerous case-studies of this in the bibliography, see for example \cite{examples}, for early examples working with twisted compactifications from the holographic point of view. The topic progressed considerably after the paper \cite{Gaiotto:2009we}. This was followed by many works studying  compactifications (twisted or with fluxes) from a purely QFT point of view. In the particular case of  compactifications of 6d to 4d systems (preserving minimal SUSY in both dimensions), we find the works \cite{examples2}-\cite{Bah:2021iaa}. For a very nice summary of these developments from a field theoretical perspective, see \cite{Razamat:2022gpm}.

 In this paper, we present an interesting example of flow across dimensions involving a twisted compactification.
In particular, we start from an infinite family of six-dimensional ${\cal N}=(1,0)$
SCFTs and compactify it on a  two manifold of constant curvature. The end-point of the flow is an infinite family of strongly coupled four dimensional ${\cal N}=1$ SCFTs (and possibly gapped QFTs, that we leave for future studies). The holographic study of the family of 4d SCFTs occupies an important part of this work, calculating observables that characterise it.

In more detail, the contents of the paper are distributed as follows.
\\
In Section \ref{sectiongeometry}, we construct a new infinite family of Massive Type IIA backgrounds that represent the flow between a family of 
six-dimensional ${\cal N}=(1,0)$ SCFTs and four dimensional ${\cal N}=1$ SCFTs. These flows are new backgrounds, not present in the bibliography. The case of gapped four dimensional systems leads to singular backgrounds, hence we leave it to future study. The charges of the brane system are discussed, with emphasis on the effects of the twisted-compactification.

 In Section \ref{QFTsection}, we present calculations of the holographic central charge in these supergravity backgrounds (the free energy of the dual CFT). These are calculations at the AdS$_5$ fixed point and along the flow. We also present a monotonic quantity interpolating between the conformal points at low and high energies. After this, based on the branes charges discussed in Section \ref{sectiongeometry},
we give a {\it phenomenological proposal} for a suitable quiver capturing the low energy dynamics. These 4d quiver QFTs are proposed to reach a conformal point at low energies, their strongly coupled dynamics being described by the infinite family of Massive Type IIA backgrounds with an AdS$_5$ factor (discussed in Section \ref{sectiongeometry}).  We emphasise on the heuristic character of this proposal. Indeed, whilst the beta functions and R-symmetry anomalies of the proposed QFT are cancelled and the scaling of the free energy with the quiver parameters (rank of gauge groups and number of nodes) matches the holographic result, the precise coefficient of the free energy does not exactly match the one computed in the holographic dual. Hence the proposed quiver is only a first step towards the correct field theory dual to our infinite family of geometries. We discuss possible improvements in the conclusions and Appendices.


In Section \ref{concl}, we summarise,  present conclusions and propose some ideas for further research.
Three very intensive appendices complement the presentation. The reader wishing to work on these topics should  benefit from reading them in detail.

\section{Supergravity backgrounds}\label{sectiongeometry}

We start this section by describing an infinite family of supergravity solutions, the analysis of which, is the main subject of the rest of this paper. This is a family  of Massive Type IIA backgrounds,  preserving four supersymmetries (${\cal N}$=1 in four dimensional notation). The construction of these backgrounds is described in great detail in Appendix \ref{appendixone}.  From a quantum field theoretical perspective, these backgrounds are dual to  twisted compactifications of six dimensional ${\cal N}=(1,0)$ SCFTs at the origin of their tensor branch. We discuss this in more detail in Section \ref{QFTsection}.

Let us present the family of backgrounds in Massive Type IIA. These are written in terms of coordinates, parameters and functions,
\begin{eqnarray}
& & \text{Coordinates:}~(t,x_1,x_2,x_3, r, \theta_1,\phi_1, z, \theta_2,\phi_2).~~\text{Parameters:}~ (\Psi_0, k).\label{funciones}\\
& & 
\text{Functions:}~\alpha(z), f(r), h(r), X(r)= e^{\frac{2}{5}\Phi(r)}, \omega(r,z)= \left( \frac{\alpha'(z)^2  - 2 \alpha(z)\alpha''(z)  X(r)^5   }{\alpha'(z)^2 -2 \alpha(z)\alpha''(z) } \right) .\nonumber
\end{eqnarray}
 The equations constraining these functions are written below. In  terms of these coordinates and functions,  the string-frame spacetime metric reads
    \begin{equation}\label{metric-z}
    \begin{aligned}
        ds^{2}_{st} &= 2\pi\sqrt{2}\sqrt{-\frac{\alpha(z)}{\alpha''(z)}}\,
        X(r)^{-\frac{1}{2}}e^{-\frac{4\Phi(r)}{5}}\left[e^{2f(r)}dx^{2}_{3,1}
        + dr^{2} + e^{2h(r)}\left( d\theta^{2}_{1} + \frac{1}{k}\sin^{2}(\sqrt{k}\theta_{1})d\phi^{2}_{1}\right)\right]\\
        & + X(r)^{5/2}\left[ \pi\sqrt{2}\sqrt{-\frac{\alpha''(z)}{\alpha(z)}}dz^{2} 
        +\frac{\sqrt{2}\,\pi}{\omega(r,z)} \frac{\sqrt{-\alpha^{3}(z)\alpha''(z)}}{2\alpha(z)\alpha''(z)-\alpha'^{2}}  \left(d\theta^{2}_{2}+\sin^{2}(\theta_{2})
        \left( d\phi_{2} - \frac{1}{k}\cos(\sqrt{k}\theta_{1})d\phi_{1}\right)^{2}\right)\right].
    \end{aligned}
    \end{equation}

    
The Neveu-Schwarz  ($B_2,\Psi$) and Ramond ($F_0, F_2, F_4$) background fields are,
  \begin{eqnarray}\label{NSRRforms}
 & & B_{2} = \left(\frac{\pi}{\omega(r,z)}\frac{\alpha(z)\alpha'(z)}{\alpha'(z)^{2}-2\alpha(z)\alpha''(z)}\sin(\theta_{2})d\theta_{2} -\pi\cos(\theta_{2})dz \right)\wedge\left( d\phi_{2} -\frac{1}{k}\cos(\sqrt{k}\theta_{1})d\phi_{1} \right),\nonumber\\
 & &  e^{4\Psi(r,z) }= \frac{X^5(r)}{\omega^2(r,z)} \left( \frac{-\alpha(z)}{\alpha''(z)}\right)^3\left( \frac{e^{2\Psi_0}}{ \alpha'(z)^{2}-2\alpha(z)\alpha''(z)     }\right)^2,\nonumber\\
 & &  F_0= 2^{\frac{1}{4}} \frac{e^{-\Psi_{0}}}{\sqrt{\pi}} \alpha'''(z) ,\label{eqF0}\\
 & & F_{2} = 2^{\frac{1}{4}}\sqrt{\pi}e^{-\Psi_{0}}\alpha''(z) \left[ \cos(\theta_{2})\Vol(\Sigma_{k}) -\Vol(S^{2}_{c}) \right] 
        + F_{0}\frac{\pi}{\omega(r,z)} \frac{\alpha(z)\alpha'(z)}{\alpha'(z)^{2}-2\alpha(z)\alpha''(z)}\, \Vol(S^{2}_{c}),\nonumber\\
& & F_{4} = \left(\frac{2^{\frac{1}{4}   } \pi^{\frac{3}{2}  } e^{-\Psi_{0}}   }{\omega(r,z)} \right) \left(\frac{\alpha(z)\alpha'(z)\alpha''(z)}{\alpha'(z)^{2}-2\alpha(z)\alpha''(z)}\right)  \cos(\theta_{2})\, \Vol(\Sigma_{k})\wedge \Vol(S^{2}) \nonumber\\
& &~~~~~+ 2^{\frac{1}{4}}\pi^{\frac{3}{2}}e^{-\Psi_{0}}\alpha''(z)  \sin^{2}(\theta_{2})\, dz\wedge d\phi_{2}\wedge \Vol(\Sigma_{k}).\nonumber
\end{eqnarray}
We have defined the volume elements,
\begin{eqnarray}
& &         \Vol(S^{2}) = \sin(\theta_{2})d\theta_{2} \wedge d\phi_{2},~~
        \Vol(S^{2}_{c}) = \sin(\theta_{2})d\theta_{2} \wedge\left( d\phi_{2}-\frac{1}{k}\cos(\sqrt{k}\theta_{1})d\phi_{1}\right),\nonumber\\
 & &    \Vol(\Sigma_k)=   \frac{\sin\left(\sqrt{k}\theta_1\right)}{\sqrt{k}} d\theta_1\wedge d\phi_1.\label{volumeforms}   
    \end{eqnarray}
%
%
    The functions $f(r), h(r),\Phi(r)$ must satisfy  first order (BPS) ordinary differential equations. Denoting the derivative respect to the coordinate $r$ with a dot, they read,
       \begin{eqnarray}
& &         \dot{f} = \pm \frac{m}{2}\,e^{-2\Phi},~~~\dot{h} = \pm \frac{1}{2}\left( \frac{1}{k}e^{-2h}+m\,e^{-2\Phi} \right),\nonumber\\
& & \dot{\Phi} = \pm\left(-1+\frac{1}{4k}e^{-2h} + m\,e^{-2\Phi}\right).\label{BPSp}
    \end{eqnarray}
We  choose the positive sign from now on. The derivation of eqs.(\ref{BPSp}) and the origin of the parameter $m$ are explained in Appendix \ref{appendixone}.   
The remaining BPS equation for $\alpha(z)$ is already written in eq.(\ref{eqF0}). In fact, the mass-parameter of massive Type IIA  $F_0$, that should be constant by pieces for an interpretation in terms of localised D8 branes dictates that $\alpha'''(z)$ must be piece-wise constant. It is in the many possible choices for a piece-wise constant $F_0$ that the  family of backgrounds is originated. More on this in Section \ref{sectionpage} below.
    
The configurations in eqs.(\ref{metric-z})-(\ref{BPSp}), are new solutions to the equations of motion of Massive IIA. In string frame these read,
    \begin{eqnarray}
 &  &\frac{1}{4}R+\nabla^{2} \Psi -(\nabla \Psi)^{2}-\frac{1}{8}H_{3}^{2}=0,\nonumber\\
     & & d F_{p} + H_{3}\wedge *F_{p-2} = 0,~~~d(e^{-2\Psi}* H_{3}) -\left(F_{0} *F_{2}+F_{2}\wedge *F_{4}+F_{4}\wedge F_{4}\right)=0,\nonumber\\
    &    & R_{MN}+2\nabla_{M}\nabla_{N} \Psi -\frac{1}{2}(H^{2}_{3})_{MN}-\frac{1}{4}e^{2\Psi}\sum_{p} (F^{2}_{p})_{MN}=0 .\label{eqs-massive-iia}
    \end{eqnarray}
 In eq. (\ref{eqs-massive-iia}) $p=2,4,6,8,10$, and
    \begin{equation}
        (F^{2}_{p})_{MN} = \frac{1}{(p-1)!}F_{M}^{\phantom{M}N_{1}...N_{p-1}}F_{N N_{1}...N_{p-1}},\;\;\; (H^{2}_{3})_{MN}= \frac{1}{2}H_{M}^{\phantom{M}N_{1}N_{2}}H_{N N_{1}N_{2}}.\nonumber
    \end{equation}
Regarding the volume form $\vol(\Sigma_k)$ in eq.(\ref{volumeforms}),  for the  allowed values for the parameter $k$, namely  $k=(1,-1)$, $\Sigma_k$ is describing a two-sphere or a hyperbolic plane. The case $k=0$, corresponding to a torus, is slightly more subtle and will be briefly addressed below.

This concludes the presentation of the backgrounds.
To interpret these in terms of branes,  we calculate the Page charges (quantised and gauge-variant) associated with this family of SUSY solutions. 

\subsection{Page fluxes and charges}\label{sectionpage}
The Page fluxes $\hat{F}_p$, defined as a polyform $\hat{F}= e^{-B_2} \wedge F$ are quantised. This implies a certain form for some of the functions in the background, as we discuss below.

 The Page fluxes are  gauge-variant. They do change under a gauge transformation of $B_2$. We use this in our favour, performing a particular transformation that makes explicit the quantised charges present and the role of the function $\alpha(z)$.
To write the Page fluxes in a concise fashion, it proves useful to define a one form $\Theta_1$ and its exterior derivative,
 \begin{equation}
        \Theta_{1} = -\cos(\theta_{2})\left( d\phi_{2} - \frac{1}{k}\cos(\sqrt{k}\theta_{1})d\phi_{1} \right),~~~
        d\Theta_{1} = \Vol(S^{2}_{c})-\cos(\theta_{2})\Vol(\Sigma_{k}).\label{Theta1}
    \end{equation}
It is also convenient to change the $B_2$-field by a large gauge transformation  (this has no effect on $H_3= dB_2$). Below, we explain the purpose of such transformation,
\begin{eqnarray}
& & B_{2,new}= B_{2,old} - \pi d\Big[ (z-\Delta) \Theta_1\Big],~~~\Delta~\text{is a constant},\label{b2new}\\
& & B_{2,new}= \left(\frac{\pi}{\omega(r,z)}\frac{\alpha(z)\alpha'(z)}{\alpha'(z)^{2}-2\alpha(z)\alpha''(z)}\Vol(S^{2}_{c}) - \pi (z-\Delta)d\Theta_1\right),\nonumber\\
& & H_{3}=\pi\left[ -\frac{1}{\omega^{2}}\frac{\alpha(z)\alpha'(z)}{\alpha'(z)^{2}-2\alpha(z)\alpha''(z)}\frac{\partial\omega}{\partial X}X'(r)dr + \frac{d}{dz}\left(\frac{1}{\omega}\frac{\alpha(z)\alpha'(z)}{\alpha'(z)^{2}-2\alpha(z)\alpha''(z)} \right)dz \right]\wedge \Vol(S^{2}_{c})\nonumber\\
 &  &\phantom{=} - \pi\left(\frac{1}{\omega}\frac{\alpha(z)\alpha'(z)}{\alpha'(z)^{2}-2\alpha(z)\alpha''(z)}\right) \sin(\theta_{2})d\theta_{2} \wedge \Vol(\Sigma_{k})
        -\pi\, dz \wedge \Vol(S^{2}_{c})+\pi\cos(\theta_{2})dz\wedge \Vol(\Sigma_{k}).\nonumber
\end{eqnarray}
With this new-$B_2$ we compute the Page flux $\hat{F}_2$ and obtain,
\begin{eqnarray}
& & \hat{F}_2= F_2- B_2 F_0=   -2^{\frac{1}{4}}\sqrt{\pi}e^{-\psi_{0}} (\alpha'' - (z-\Delta)\alpha''')d\Theta_{1}    . \label{F2page} 
\end{eqnarray}
Similarly we calculate the $\hat{F}_4$ Page flux,
\begin{eqnarray}
& & \hat{F}_4= F_4- B_2\wedge F_2+\frac{1}{2} B_2\wedge B_2 F_0,\label{f4page}\\
&  &
\hat{F}_{4} ={  2^{\frac{1}{4}}\pi^{\frac{3}{2}}e^{-\Psi_{0}}\alpha''(z) \sin^2\theta_2 \, dz\wedge d\phi_{2} \wedge \Vol(\Sigma_{k})  }\nonumber\\
& & ~~~~~{+  2^{1/4} \pi^{3/2} e^{-\Psi_0} (z-\Delta) \left(  2\alpha''(z)  - \alpha'''(z)(z-\Delta)     \right)  \cos\theta_2 \Vol(S^2_c)\wedge \Vol(\Sigma_k).} \nonumber
\end{eqnarray}
Finally, for $F_0$ we have the same as in eq.(\ref{eqF0}),
$
\hat{F}_0= 2^{\frac{1}{4}} \frac{e^{-\Psi_{0}}}{\sqrt{\pi}} \alpha'''(z)$.
We now calculate the charges associated with these Page fluxes, and impose their quantisation.
\subsubsection{Page charges}
To calculate the charges,  we need to  compute the integrals (as in the rest of the paper we set $g_s=\alpha'=1$), 
 \begin{equation}
      Q_{NS5}=\frac{1}{4\pi^2}\int_{M_3} H_3,~~  Q_{D_p}=\frac{1}{(2\pi)^{7-p}}\int_{M_{8-p}}\widehat{F}_{8-p},~~~ p=4,6,8.
    \end{equation}
These integrals need to be defined over suitable cycles, some of which contain the sub-manifold $\Sigma_k$. Then, it is useful to first calculate the volume of the two-manifold $\Sigma_{k}$ defined in eq.(\ref{volumeforms}). As we stated above, in the cases $k=(1,0,-1)$ the space described is a two-sphere, a torus or a hyperbolic plane.
Its volume is calculated using Gauss-Bonnet's theorem and the fact that $\Sigma_{k}$ has curvature $R=2k$. For a genus $g$ Riemann surface, we have
 \begin{equation}
        \int_{\Sigma_{k}} d^{2}x \sqrt{g} R = 8\pi(1-g)~~\longrightarrow ~~   \int_{\Sigma_{k}} d^{2}x \sqrt{g} = \frac{4\pi}{k}(1-g).\nonumber
    \end{equation}
For  the sphere, $k=1$ and $g=0$, while for the hyperbolic plane,  $k=-1$ with  $g>1$. The case of the torus needs some care, as we have $k=0, g=1$ (and the volume is $4\pi$). We will not discuss the case of $T^2$ in what follows (except for the purpose of making an intuitive argument below). This allows us to write (for $S^2, H_2$)
    \begin{equation}
        \int_{\Sigma_{k}} d^{2}x \sqrt{g} = 4\pi |g-1|.\label{volumesigma2}
    \end{equation}
 After these preliminaries, we calculate the Page charges.
Inspecting the $H_3$-field in eq.(\ref{b2new}) we find two possible three-cycles on which the integral can be performed. These three-cycles are
\begin{equation}
{\cal M}_{1}=(\theta_{1},\phi_{1},z)\big|_{\theta_{2}=0},\;\;\; \text{and} ~~{\cal M}_{2}=(\theta_{2},\phi_{2},z)\big|_{r\rightarrow+\infty}.
\end{equation}
We find that there are two sets of NS-five branes. Their total numbers being,
    \begin{eqnarray}
& &        N^{(1)}_{NS5} = \frac{1}{4\pi^{2}} \int_{{\cal M}_{1}} H_{3} = \frac{1}{4\pi}\Vol(\Sigma_{k})P= |g-1| P,\nonumber\\
 & &       N^{(2)}_{NS5} = \frac{1}{4\pi^{2}} \int_{{\cal M}_{2}} H_{3} = P.\label{cargasns}
 \end{eqnarray}
To calculate $ N^{(2)}_{NS5}$ a boundary condition $\alpha(z=0)= \alpha(z=P)=0$ has been imposed, more on this below. We have also taken the orientation of the manifolds  such that all  the charges are positive. Importantly, we have set the range of the $z$-coordinate to be $z\in [0,P]$, with $P$ an integer.
   
For the charges of D6-branes, we have a pair of two-cycles on which to integrate the Page flux of eq.(\ref{F2page}),
\begin{equation}
{\cal M}_{3}=(\theta_{1},\phi_{1})\big|_{\theta_{2}=0},\;\;\; \text{and}\;\;\; {\cal M}_{4}=(\theta_{2},\phi_{2}).\nonumber
\end{equation}
  Performing the integrals, we find,
 \begin{eqnarray}
& &         N^{(1)}_{D6} = \frac{1}{2\pi} \int_{{\cal M}_{3}} \hat{F}_{2} 
        = 2^{\frac{5}{4}}\sqrt{\pi}e^{-\Psi_{0}}|g-1|(\alpha'' - (z-\Delta)\alpha''') ,\nonumber\\
 & &        N^{(2)}_{D6} = \frac{1}{2\pi} \int_{{\cal M}_{4}} \hat{F}_{2} 
        = 2^{\frac{5}{4}}\sqrt{\pi}e^{-\Psi_{0}} (\alpha'' - (z-\Delta)\alpha''').\label{charged6}
 \end{eqnarray}
For the D8 branes we use eq.(\ref{eqF0}) and find,
    \begin{equation}
        N_{D8} = 2\pi\int dz F_{0}' 
        = 2^{\frac{5}{4}}\sqrt{\pi}e^{-\Psi_{0}}\int dz\, \alpha^{(4)}.\label{charged8}
    \end{equation}
{
Finally, for the D4 branes, the four-cycle is $  {\cal M}_5=(\theta_2,\phi_2,\theta_1,\phi_1)$, at constant values of $r,z$. Calculating
\begin{equation}
Q_{D4}= \frac{1}{8\pi^3} \int_{{\cal M}_5} \hat{F}_4=0.\nonumber
\end{equation}
In other words, there is no charge of D4 branes in the system\footnote{One might wonder about computing $Q_{D4}$ integrating $\hat{F}_4$ over the manifold ${\cal M}_6=[z,\theta_1,\phi_1, \phi_2]_{\theta_2=\frac{\pi}{2}}$. This is not a well defined four-cycle, as it has a boundary.      }.}

Inspecting the charges in eqs.(\ref{cargasns}), (\ref{charged6}) and (\ref{charged8}), suggests to set $e^{\Psi_{0}}=2^{\frac{5}{4}}\sqrt{\pi}$. Imposing quantisation, it is clear from eqs. (\ref{charged6})
that the function $\alpha''(z)$ must be a linear function with integer coefficients $N_l$. In fact, we can divide the range of the $z$-coordinate in intervals of unit size. In each interval, $\alpha''(z)$ should be a linear function. The charge of D8 branes in eq.(\ref{charged8}) suggests that $\alpha''(z)$ should be piecewise linear and continuous (with integer coefficients), the third derivative piecewise constant (integer and  generically discontinuous), whilst the fourth-derivative a sum of delta functions with integer coefficients. In other words, if we choose for $\alpha''(z)$,
    \begin{equation}
    \alpha''(z) = 
       \begin{cases}
       N_{1}z &, 0\leq z < 1\\
       N_{1} + (N_{2}-N_{1})(z-1) &, 1\leq z < 2\\
       \vdots\\
       N_{l} + (N_{l+1}-N_{l})(z-l) &, l\leq z < l+1\\
       \vdots\\
       N_{P-1}(P-z)  &, (P-1)\leq z < P.\\\label{alphasecond}
     \end{cases}
    \end{equation}
This implies,
    \begin{equation}
     \alpha'''(z) = 
       \begin{cases}
       N_{1} &, 0\leq z < 1\\
       N_{2}-N_{1} &, 1\leq z < 2\\
       \vdots\\
       N_{l+1}-N_{l} &, l\leq z < l+1\\
       \vdots\\
       -N_{P-1}  &, (P-1)\leq z < P,\\
     \end{cases}
    \end{equation}
and
    \begin{equation}
       \alpha^{(4)}(z) = \sum^{P-1}_{l=1} \left( 2N_{l}-N_{l+1}-N_{l-1} \right)\delta(z-l).\label{alpha4}
    \end{equation}
At this point we choose a convenient coefficient $\Delta$ in the large gauge transformation of $B_2$-- see eq.(\ref{b2new}) . In fact, choosing $\Delta=l$ for  $z\in [l,l+1]$ we have in each interval 
\begin{equation}
 (\alpha'' - (z-l)\alpha''') =N_l.\nonumber
 \end{equation}
%
In  summary, we have two kinds of NS-five branes, their total number is given in eqs.(\ref{cargasns}). In the interval $z\in[l,l+1]$ we have  two types of D6 brane charges and  one kind of D8 charge given by (in each interval),
    \begin{eqnarray}
       & &N^{(1)}_{D6}[l,l+1] \,=  |g-1|N_{l} ,\;\;\; 
        N^{(2)}_{D6}[l,l+1] \,=  N_{l},\nonumber\\
 &       &N_{D8}[l,l+1] \,= 2N_{l}-N_{l+1}-N_{l-1} .\label{chargesinterval}
    \end{eqnarray}
    Though it does not feature in the calculation of Page charges,  the function $\alpha(z)$ is obtained after two integrations
    of eq.(\ref{alphasecond}). The integration constants must be chosen such that $\alpha(z)$ is a piecewise continuous cubic function with continuous derivative $\alpha'(z)$. To avoid singular behaviours (not associated with the presence of localised D8 branes), it must satisfy $\alpha(0)=\alpha(P)=0$.
     
This is a good point to discuss the physical effect of the large gauge transformation on the $B_2$-field-- see eq.(\ref{b2new}). Being a gauge transformation, it does not affect the Physics of our system, but it makes the counting of charges more transparent. Had we not performed it and calculated Page charges with the $B_{2,old}$, we would have obtained a combination between charges of D6 branes induced on the D8 branes and those of 'actual' D6 branes. The large gauge transformation separates these, making the counting clearer.

To better understand these systems, it is a good (and intuitive) guide to go back to the case of the torus. In this case, we are compactifying a six dimensional ${\cal N}=(1,0)$ SCFT on $T^2$ without any flux that breaks SUSY. In other words, we would end with an ${\cal N}=2$ four dimensional SCFT. For more on this perspective see, for example \cite{Baume:2021qho}.  What follows in the next paragraph is an {\it intuitive} argument.

The compactification  on $T^2$ (which sets $k=0$) should be handled with some care. For example, the 'twisting' in the one-form $\Theta_1$ is absent, leading to $\Theta_1=-\cos\theta_2 d\phi_2$. Also,  in the BPS eqs.(\ref{BPSp}), the terms $\frac{e^{-2h} }{k} $ are absent. This gives  $2 \Phi(r)=0$ with $X(r)=1$ (for the parameter $m=1$) and $2f(r)=2h(r)= r$. This leads to a background  metric---see eq.(\ref{metric-z})-- of the form AdS$_7\times S^2\times R_z$. For this case the genus is $g=1$ and the charges in eqs.(\ref{cargasns}) and (\ref{chargesinterval}) indicate the presence of only one type of NS-five branes and D6 branes, together with D8 branes. The reader will recognise that this is the compactification on $T^2$ of the backgrounds in \cite{sixd}-\cite{Nunez:2018ags}. 

Intuitively, this is what our background in eqs.(\ref{metric-z})-(\ref{BPSp}) is describing for large values of the $r$-coordinate. Of course, since we compactify on a curved two manifold (either $S^2$ or $H_2$), the twisting needs to be performed even at high energies (large values of the $r$-coordinate). The system preserves four supercharges all along the flow. The twisted compactification of the six dimensional system of NS-D6-D8 induces  new sets of NS  and D6 branes. We identify these new branes as those whose charge comes with the $(g-1)$-factor in front-- see eqs.(\ref{cargasns}) and (\ref{charged6}).

Let us be more precise about the compactifications on $S^2$ and $H_2$. 
\subsection{The cases of $H_2$ and $S^2$ compactifications}
\begin{figure}[h!]
        \centering
        \includegraphics[width=0.5\linewidth]{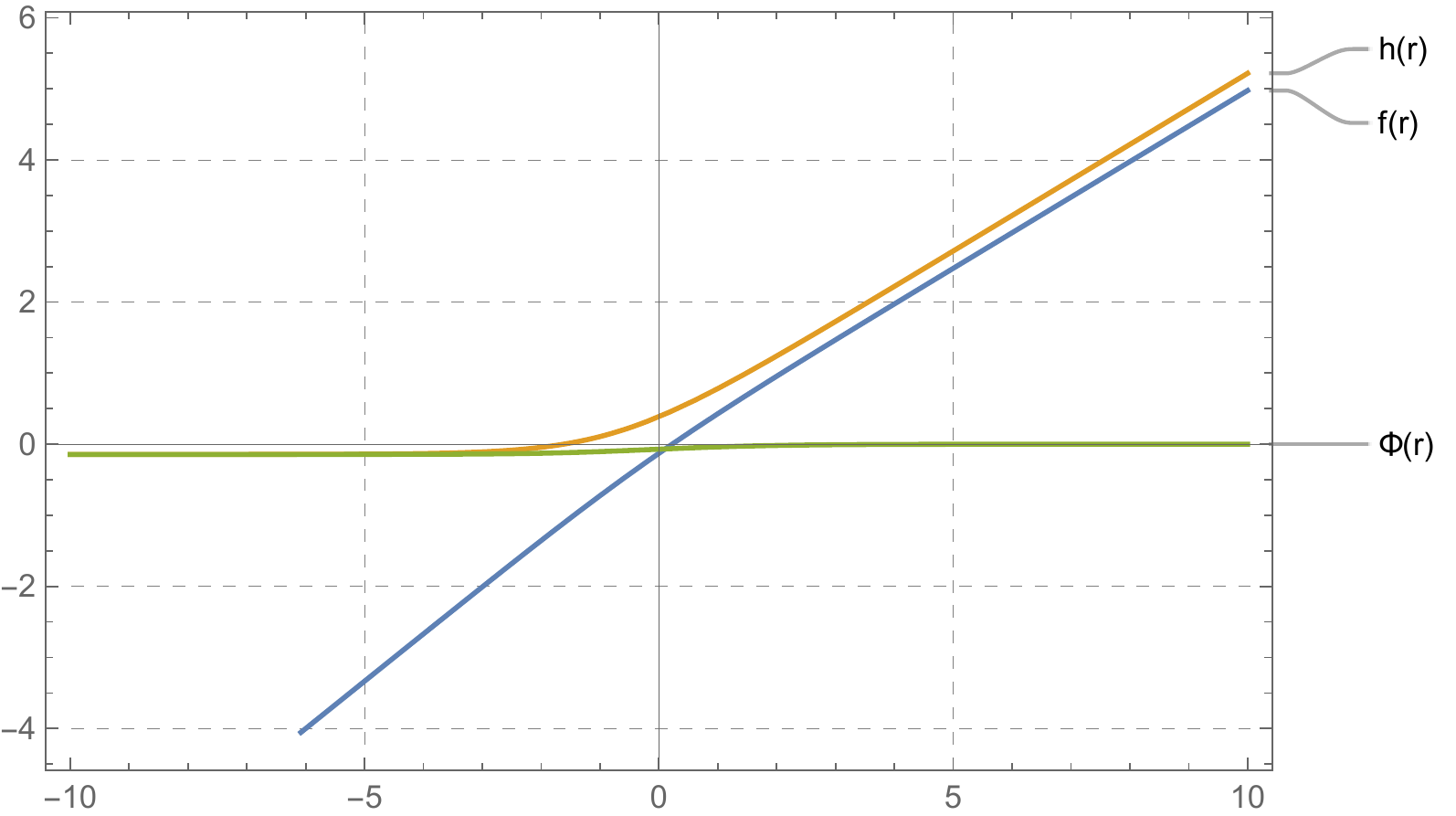}
        \caption{The plot for the case of the $H_2$ compactification ($k=-1$) of the functions $f(r), h(r), \Phi(r)$. The UV asymptotics at $r\to +\infty$ and the IR at $r\to -\infty$. We clearly see the attainment of both IR and UV fixed points. }
    \label{Functionspdf}
    \end{figure}

In both  cases ($k=1$ or $k=-1$), we can solve two of the equations in (\ref{BPSp}), finding
\begin{equation}
e^{-2\Phi}= \frac{2 k \dot{h} - e^{-2h}}{k m},~~~f= \frac{m}{2} \int e^{-2\Phi} dr,
\end{equation}
whilst $e^{2h(r)}={\cal G}(r)$ must satisfy a nonlinear second order differential equation,
\begin{equation}
\ddot{ {\cal G} } +\frac{( \dot{ { \cal G } } )^2}{ {\cal G} } -\frac{5 \dot{ {\cal G}  } }{2 k {\cal G}}- 2 \dot{{\cal G}} +\frac{3}{2{\cal G}}+\frac{2}{k}=0.
\end{equation}
The numerical resolution of this system is studied in Appendix \ref{appendixtwo}. Let us gain some understanding by discussing asymptotic solutions.
For large values of the coordinate $r$ we find an asymptotic solution,
\begin{equation}
e^{2h(r)}\sim e^r,\;\;\; e^{2f(r)}\sim e^{2f_0 +r},\;\;\; e^{2\Phi}\sim \frac{km}{k-e^{-r}}\sim m.\label{large-r-asymp}
\end{equation}
In the case of the compactification on the hyperbolic plane ($k=-1$), we find an exact, fixed-point  solution. This is  the same solution found in \cite{Bah:2017wxp}, after conventions are matched.
\begin{equation}
e^{2f(r)}= e^{\frac{2}{3} r},\;\;\; e^{2h(r)}=\frac{3}{4},\;\;\; e^{2\Phi(r)}= \frac{3m}{4}.\label{fixedpoint}
\end{equation}
As is usual, the $r$-coordinate plays the role of energy-coordinate. The solution in eq.(\ref{large-r-asymp}) asymptotes to $\widehat{\text{AdS}}_7\times S^2_c \times R_z$ when replaced in eqs.(\ref{metric-z})-(\ref{eqF0}), describing  a six dimensional SCFT formulated on $R^{1,3}\times H_2$. The $\widehat{\text{AdS}}_7$ is written as a foliation over this six-space. On the other end,
 when the fixed point solution of eq.(\ref{fixedpoint}) is replaced in the family of backgrounds of Massive Type IIA of  eqs.(\ref{metric-z})-(\ref{eqF0}), the space time takes the form AdS$_5\times {\cal M}_5$. These  describe the 
  dual to a family of four dimensional ${\cal N}=1$ SCFTs. The numerical solution connecting  the large-$r$ asymptotics in eq.(\ref{large-r-asymp}) with the fixed-point  exact solution (\ref{fixedpoint}) is described in Appendix \ref{appendixtwo}. We present the plot for the functions $f(r), h(r), g(r)$ in Figure \ref{Functionspdf}.

 The numerical study of the case of the two-sphere ($k=1$) leads to badly singular behaviour as we decrease the $r$-coordinate. We believe that a background 
more elaborated than  that in eqs.(\ref{metric-z})-(\ref{eqF0}) could resolve this singular behaviour, but leave this for a future investigation. In what follows we focus our attention on the backgrounds describing the  twisted compactification on $H_2$ (preserving 4 SUSYs) of a six dimensional ${\cal N}=(1,0)$ 6d SCFT.

%
 
 
 We close this section emphasising that we have presented a new infinite family of backgrounds in eqs.(\ref{metric-z})-(\ref{BPSp}). Each solution in this family is labelled by each possible $\alpha(z)$. At  the IR fix point, this is a Massive IIA background with an AdS$_5$ factor preserving four supercharges.

\section{Study of the dual field theory}\label{QFTsection}
We start  this section with the holographic
calculation of the free energy of the field theory. In particular we start with the free energy of the 4d SCFT that results in the low energy limit of our compactification. After that, we study the same quantity along the flow described by our full solution. We then define a monotonic quantity that captures the flow between the low energy four dimensional SCFT and the high energy six dimensional conformal point (as suggested by the family of supergravity backgrounds). 

After these holographic calculations we use the charges calculated in Section \ref{sectiongeometry} to give an {\it heuristic} proposal for a 4d quiver that at low energies becomes conformal and captures some aspects of the fixed point AdS$_5$ backgrounds. This proposal is provisional and certainly incomplete. In fact, whilst various observables and scaling behaviours do match between supergravity and QFT calculations, the precise coefficient of the free energy does not, suggesting that some extra fields should supplement our QFT proposal.

\subsection{Free Energy and Holographic Central Charge}\label{centralsection}
One meaningful observable of SCFTs is the free energy. For the case of SCFTs in diverse dimensions this was calculated holographically. See for example the papers  \cite{Lozano:2019zvg},
\cite{Coccia:2020wtk},
 \cite{Akhond:2021ffz},
\cite{4dss}, 
 \cite{Uhlemann:2019ypp},
   \cite{Legramandi:2021uds},
 \cite{sixd}.
These calculations were checked against field theoretical computations (typically a global anomaly coefficient or a localisation calculation in matrix models). These checks of the AdS/CFT correspondence are valid in the regime of parameters for which the supergravity background is a reliable representation of the QFT dynamics. This is typically for long linear quivers $P\to\infty$ and for large rank of the gauge nodes $N_i\to\infty$.

The calculations done with the supergravity backgrounds are either computing a regularised on-shell action for a putative reduced supergravity or alternatively the calculation of the Newton constant associated with  a reduced  theory of gravity in lower dimensions.

In particular, it was shown in \cite{Macpherson:2014eza}, \cite{Bea:2015fja} that for any generic holographic background dual to a QFT in $d+1$ spacetime dimensions, with metric and dilaton given by,
 \begin{equation}
        ds^{2} = a(r,y^{i}) \left( dx^{2}_{d,1}+b(r) dr\right) + g_{ij}dy^{i} dy^{j},~~~~\Psi(r,y^i),\label{cani}
    \end{equation}
we can define quantities $V_{int}, \hat{H}$ according to
 \begin{equation}
 V_{int}= \int dy^i \sqrt{e^{-4 \Psi} a(r, y^i)^d \det[g_{int}]},~~~~\hat{H}= V_{int}^2.\label{vint}
 \end{equation}
From these we define the holographic central charge (or free energy),
    \begin{equation}
        c_{hol} = d^{d} \frac{b(r)^{\frac{d}{2}} H^{\frac{2d+1}{2}}}{G^{(10)}_{N} (H')^{d}},\label{chol}
    \end{equation}
    where $G^{(10)}_{N} = 8\pi^6$ is in our conventions, the ten-dimensional Newton constant. 

 For the case of our backgrounds in eq.(\ref{metric-z}), we compare with eq.(\ref{cani}) and  read,
    \begin{equation}
        a(r,z)=2\beta\pi\sqrt{2}\sqrt{-\frac{\alpha(z)}{\alpha''(z)}}\,
        X(r)^{-\frac{1}{2}}e^{-\frac{4\Phi(r)}{5}+2f(r)}, ~~ b(r) = e^{-2f(r)}, ~~d=3.\label{vcvc}
    \end{equation}
The metric of the internal space has determinant,
    \begin{equation}
        \text{det}(g_{\text{int}}) = \frac{16\sqrt{2} \pi^{5}}{\omega^{2}}e^{4h(r)+\Phi(r)} 
            \frac{\sqrt{-\alpha(z)^7\alpha''(z)}}{(\alpha'(z)^{2}-2\alpha(z)\alpha''(z))^{2}} \frac{\sin^{2}(\sqrt{k}\theta_{1})}{k}\sin^{2}(\theta_{2}).
    \end{equation}
This leads, after using eqs.(\ref{vint})-(\ref{chol}), to
    \begin{eqnarray}
& &         H = \hat{\mathcal{N}}^{2}  e^{6f(r)+4h(r)-4\Phi(r)},~~
        \hat{\mathcal{N}}= 16\sqrt{2} \pi^{4} e^{-2\Psi_{0}}  \int \left[-\alpha(z)\alpha''(z)\right] dz\, \Vol(\Sigma_{k})\Vol(S^{2}).\nonumber\\
        & &  c_{hol} = \frac{27 \hat{\mathcal{N} }}{8 G^{(10)_{N}}} \frac{e^{2h(r)-2\Phi(r)}}{(3 \dot{f}+2 \dot{h}-2 \dot{\Phi})^{3}}.\label{HN}
    \end{eqnarray}
After using the BPS equations in (\ref{BPSp}),  the holographic central charge reads
    \begin{equation}
        c = \frac{27\hat{ \mathcal{N }}}{G^{(10)}_{N}} \frac{e^{2h(r)-2\Phi(r)}}{\left( 
        4 + \frac{1}{k}e^{-2h(r)} + m\, e^{-2\Phi(r)}\right)^{3}}.\label{cholf}
    \end{equation}
Since this computes the central charge of the four dimensional SCFTs-- we have set $d=3$ in eq.(\ref{vcvc})--- 
we evaluate this quantity for $k=-1$, the hyperbolic plane twisted compactification of the 6d SCFTs (setting $m=1$), and find
    \begin{equation}
        c_{hol} = \frac{27\hat{\mathcal{N} }}{64\, G^{(10)}_{N}}.\label{chol4d}
    \end{equation}
Let us analyse the results in eqs.(\ref{cholf})-(\ref{chol4d}).

First, notice that the factor $\hat{{\cal N}}$ in eq.(\ref{HN}), has a part proportional to $\Big(-  \Vol(S^2)\int dz \alpha'' \alpha \Big)$. We recognise this as information coming from the UV, six dimensional SCFT. Indeed, this factor appears  when calculating the free energy of a six dimensional $(1,0)$ SCFT, see equation (2.14) in the paper \cite{Nunez:2018ags} or equation (4.10) in the first paper referred in \cite{sixd}. This is coming from the UV-part of the flow.
%
 Note that  $\hat{{\cal N}}$ in eq.(\ref{HN}), also contains a factor of $\vol\Sigma_k=4\pi |(g-1)|$. This suggests that the number of fields gets multiplied by $(g-1)$. Both these factors inform the phenomenological proposal for the QFT in Section \ref{pheno}.

Let us evaluate the holographic central charge in eq.(\ref{chol4d}) for  two examples. Whilst these examples are  not generic, they capture many aspects of the dynamics of the QFTs. We also discuss these examples (purely from a QFT perspective) in Appendix \ref{sec-analysisqft}.

\subsubsection{ Example 1}
We study the case corresponding to 
a function $\alpha(z)$ given by,
 \begin{equation}
    \alpha(z) = 
       \begin{cases}
       \frac{N}{6} (1-P^2)z +\frac{N}{6} z^3&, 0\leq z \leq (P-1)\\
       \\
       -\frac{1}{6} (2P^2-3P+1)(P-z) +\frac{N}{6} (P-1) (P-z)^3 &, (P-1)\leq z \leq P.\label{alpha-ex1}
     \end{cases}
    \end{equation}
    This function is associated with a 6d SCFT consisting of a liner quiver with gauge groups of rank $N_j=j N$, ending with a flavour group $SU(PN)$.
The  function  $ \alpha(z)$ vanishes at $z=0$ and $z=P$, is continuous at $z=(P-1)$ and the derivative $\alpha'(z)$, is continuous at the same point. The second derivative is
\begin{equation}
    \alpha''(z) = 
       \begin{cases}
       N z&, 0\leq z \leq (P-1)\\
       \\
       N(P-1) (P-z) &, (P-1)\leq z \leq P.\label{alpha''-ex1}
     \end{cases}
    \end{equation}

We now use the expressions in eqs. (\ref{HN})-(\ref{chol4d}),  the fact that $e^{2\Psi_0}= 4\sqrt{2} \pi $, $G_N= 8\pi^6$ and find,
\begin{eqnarray}
& & \hat{{\cal N}}= \frac{64}{45}\pi^5 N^2 P^5 (g-1),~~~c_{hol,ex1}= \frac{3}{40\pi}N^2 P^5 (g-1) \left(1+O(\frac{1}{P^2}) \right).\label{chol4dex1}
\end{eqnarray}
We find a scaling with the length of the quiver and the number of nodes, that is reminiscent of what occurs for this kind of quiver for the case of six dimensional SCFTs.
Let us see a second example.
\subsubsection{Example 2}

Consider the function $\alpha(z)$,
 \begin{equation}
    \alpha(z) = 
       \begin{cases}
       \frac{N}{2} (1-P)z +\frac{N}{6} z^3&, 0\leq z \leq 1\\
       \frac{N}{6} -\frac{PN}{2} z +\frac{N}{2}z^2 &, 1\leq z \leq (P-1)\\
       -\frac{N}{2} (P-1)(P-z) +\frac{N}{6}  (P-z)^3 &, (P-1)\leq z \leq P.\label{alpha-ex2}
     \end{cases}
    \end{equation}
From the perspective of the 6d SCFT, this function describes a linear quiver with $(P-1)$
gauge nodes of rank $N_j=N$ and flavour nodes $F_J= N(\delta_{J,1}+ \delta_{J, P-1})$.
The one in eq.(\ref{alpha-ex2}) is a function that vanishes at $z=0$ and $z=P$, is continuos at $z=1$ and at  $z=(P-1)$ and the derivative $\alpha'(z)$ is continuous at the same two points. The second derivative is
\begin{equation}
    \alpha''(z) = 
       \begin{cases}
       N z&, 0\leq z \leq 1\\
       N &, 1\leq z \leq (P-1)\\
       N (P-z) &, (P-1)\leq z \leq P.\label{alpha''-ex2}
     \end{cases}
    \end{equation}
 Using eqs. (\ref{HN})-(\ref{chol4d}), we calculate,
\begin{eqnarray}
c_{hol,ex2}= \frac{9}{32\pi}N^2 P^3(g-1) \left(1+O(\frac{1}{P^2}) \right).\label{chol4dex2}
\end{eqnarray}
This 4d SCFT has a scaling characteristic of a six dimensional linear quiver of the same type described above.

This result, together with the one in the first example and their respective comparison with the field theoretical calculation
in eqs.(\ref{a-cexample1})-(\ref{a-cexample2}),  inform our phenomenological proposal of Section \ref{pheno}.
Let us now discuss the $r$-dependence of the holographic central charge in eq.(\ref{HN}).
 
\subsubsection{ Energy dependence of the holographic central charge}
Let us now interpret the $r$-dependence of our free energy/holographic central charge in eq.(\ref{cholf}). Inspecting this quantity, we find at the IR fixed point the result in eq.(\ref{chol4d}). We associate with this the free energy of a family of four dimensional  SCFTs preserving four Poincare supercharges. The family of SCFTs in 4d is labelled by the different functions $\alpha(z)$ that we choose as input, as in the examples above. On the other end, at high energies, we find using eq.(\ref{large-r-asymp}), that the quantity in eq.(\ref{cholf}) diverges. The plot of Figure \ref{figuran2} with the numerical solution found in Appendix \ref{appendixtwo} shows this.

    \begin{figure}[h!]
        \centering
        \includegraphics[width=0.5\linewidth]{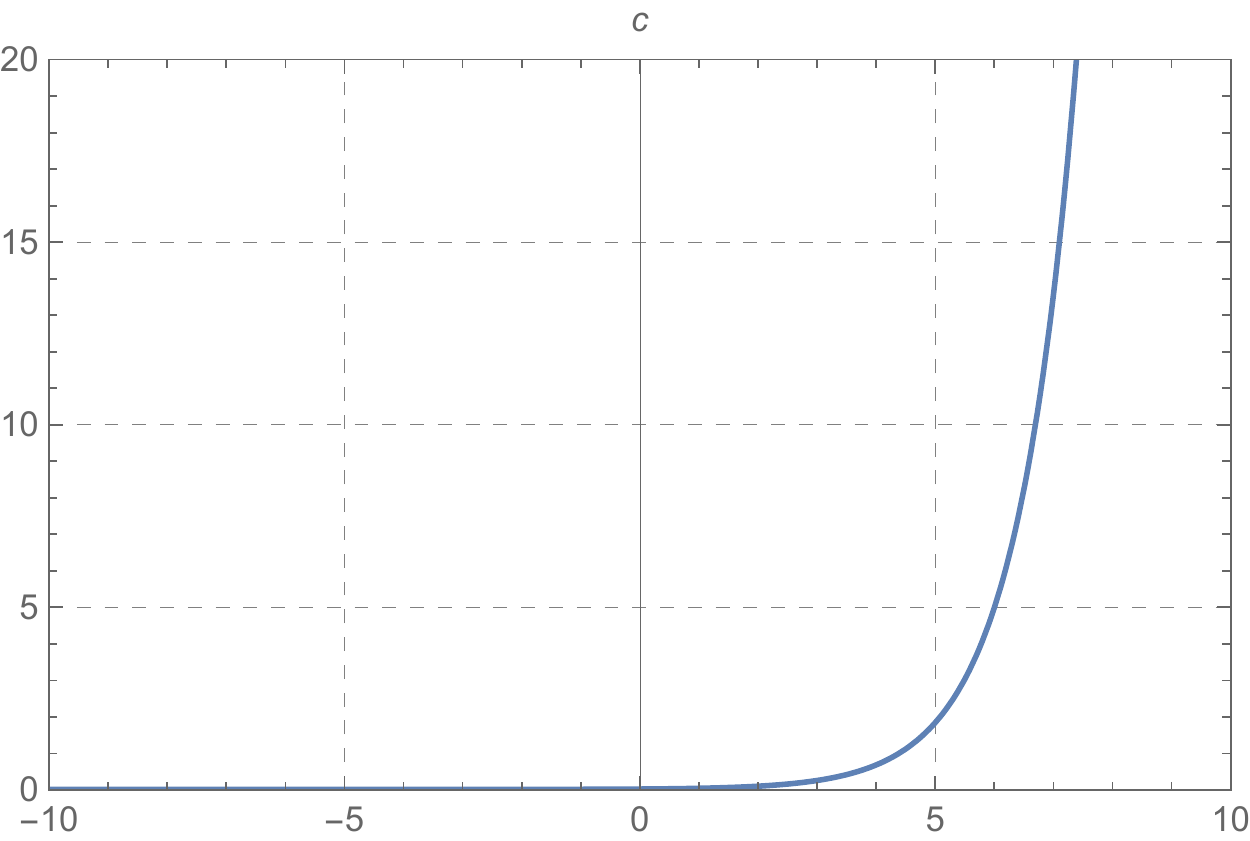}
        \caption{Central Charge calculated for the numerical solutions of Figure \ref{Functionspdf}. Here we see that the central charge is divergent in the UV.}\label{figuran2}
    \end{figure}

The divergence of this monotonically increasing (towards the UV) free energy is understandable. Massive fields that originated in the wrapping of both D6's and NS-five branes on the surface $\Sigma_k=H_2$ generically have a mass inversely proportional to the size of $\Sigma_k$ and are frozen at low energies (of course, there are also massless fields present in the 4d SCFT). When flowing towards the UV these massive fields become active. The number of these Kaluza-Klein modes grows with energy and leads to the divergence of $c_{hol}$. Somewhat, the quantity defined in eqs.(\ref{vcvc})-(\ref{chol4d}) for a {\it four} dimensional QFT (we have set $d=3$)  is not able to recognise that the system is approximately approaching a six dimensional SCFT.

Similar arguments were expressed by the authors of \cite{GonzalezLezcano:2022mcd}.  In that paper the authors 
define a quantity $c_{monotonic}$ using the Entanglement Entropy. The quantity $c_{monotonic}$ in \cite{GonzalezLezcano:2022mcd} and our eq.(\ref{cholf}), are  related by a proportionality factor when studied in our family of backgrounds.

This calls for  a quantity that is actually able to detect both fixed points (this was also one of the objectives in \cite{GonzalezLezcano:2022mcd}). A quantity called $c_{flow}$  in \cite{Bea:2015fja}, \cite{Legramandi:2021aqv} can play this role. This quantity is monotonic, but decreases towards the UV fixed point. We study this  below.

\subsection{Flow central charge}
Let us now define a quantity that is capable of detecting both the UV and IR fixed points of our flow across dimensions. With this in mind, we write a generic holographic metric dual to a QFT defined on anisotropic spacetime--see \cite{Bea:2015fja}, \cite{Legramandi:2021aqv} for details, 
   \begin{equation}
        ds^{2} = - a_{0} dt^{2}_{0} + \sum^{d-p}_{i=1} a_{i}dx^{2}_{i} + a_{d-p+1}ds^{2}_{p} + \prod^{d-p}_{i=1} (a_{i}\,a^{p}_{d-p+1})^{\frac{1}{d}}\, \tilde{b}\, dr^{2} + h_{ab}dy^{a}dy^{b}.
    \end{equation}
where $t$ and $x_{i}$ are coordinates of the macroscopic space, $p$ is the dimension of the compactification manifold and $ds^{2}_{p}$ its line element. The metric and the coordinates of the internal manifold are denoted by  $h_{ab}$ and $y^{a}$ respectively. The dilaton is $\Psi(r,y^a)$.
We define the quantities,
   \begin{eqnarray}
& &         G_{ij}dX^{i}dX^{j} = \sum^{d-p}_{i=1} a_{i}dx^{2}_{i} + a_{d-p+1}ds^{2}_{p} + h_{ab}dy^{a}dy^{b}, \label{cflow}\\
& &
        \tilde{H} = \left( \int \prod^{8-d}_{a=1} dX^{a} \sqrt{e^{-4\Psi}  \text{det}(G_{ij})} \right)^{2},\;\;\;
        c_{flow} = d^{d} \frac{\tilde{b}^{\frac{d}{2}} \hat{H}^{\frac{2d+1}{2}}}{G^{(10)}_{N} (\hat{H}')^{d}}.\nonumber 
    \end{eqnarray}
    Specialising these quantities in our metric of eq.(\ref{metric-z}), we have
    \begin{eqnarray}
    & & d=5, \;\;\;\;p=2,\;\;\;\;   a_{0} = a_{1} = a_{2} = a_{3} 
        =  2\pi\sqrt{2}\sqrt{-\frac{\alpha(z)}{\alpha''(z)}}\,
        X(r)^{-\frac{1}{2}}e^{-\frac{4\Phi(r)}{5}+2f(r)},\nonumber\\
        & & a_{4} = a_{0}\,e^{2h(r)-2f(r)}, \;\;\;  \tilde{b} = e^{-\frac{6}{5}f(r)-\frac{4}{5}h(r)},
 \nonumber\\
        & &   \tilde{H} = \hat{\mathcal{N}}^{2} e^{6f(r)+4h(r)-4\Phi(r)},\;\;\;   \hat{\mathcal{N}} = 16\sqrt{2} \pi^{4} e^{-2\Psi_{0}} \int(-\alpha(z)\alpha''(z)) dz\, \Vol(\Sigma_{k})\Vol(S^{2}).\nonumber
\end{eqnarray}
Using eq.(\ref{cflow}) we compute,
\begin{eqnarray}
& &        c_{flow} = \frac{27 \hat{\mathcal{N}} }{32 G^{(10)}_{N}} \frac{e^{-2\Phi(r)}}{(3\dot{f}+2\dot{h}-2\dot{\Phi})^{5}},~~ \text{use BPS in eq.(\ref{BPSp}) to find},\nonumber\\
& &
        c_{flow} = \frac{27 \hat{\mathcal{N}}}{G^{(10)}_{N}} \frac{e^{-2\Phi(r)}}{\left( 4 + \frac{1}{k}e^{-2h(r)} + m\, e^{-2\Phi(r)} \right)^{5}}.\label{cflowhere}
    \end{eqnarray}
Evaluating in the IR fixed point of eq.(\ref{fixedpoint}) and in the asymptotic UV-expansion of eq.(\ref{large-r-asymp}), leads to
    \begin{equation}
        c_{flow, IR} = \frac{9\hat{\mathcal{N}}}{256 \, G^{(10)}_{N}},\;\;\; c_{flow, UV}= \frac{27\hat{\mathcal{N} } }{5^5 \, G^{(10)}_{N} }.
    \end{equation}
We can evaluate this quantity numerically on the flow solutions discussed in Appendix \ref{appendixtwo}. The plot in Figure \ref{centralflow} shows this.
    \begin{figure}[h!]
        \centering
        \includegraphics[width=0.5\linewidth]{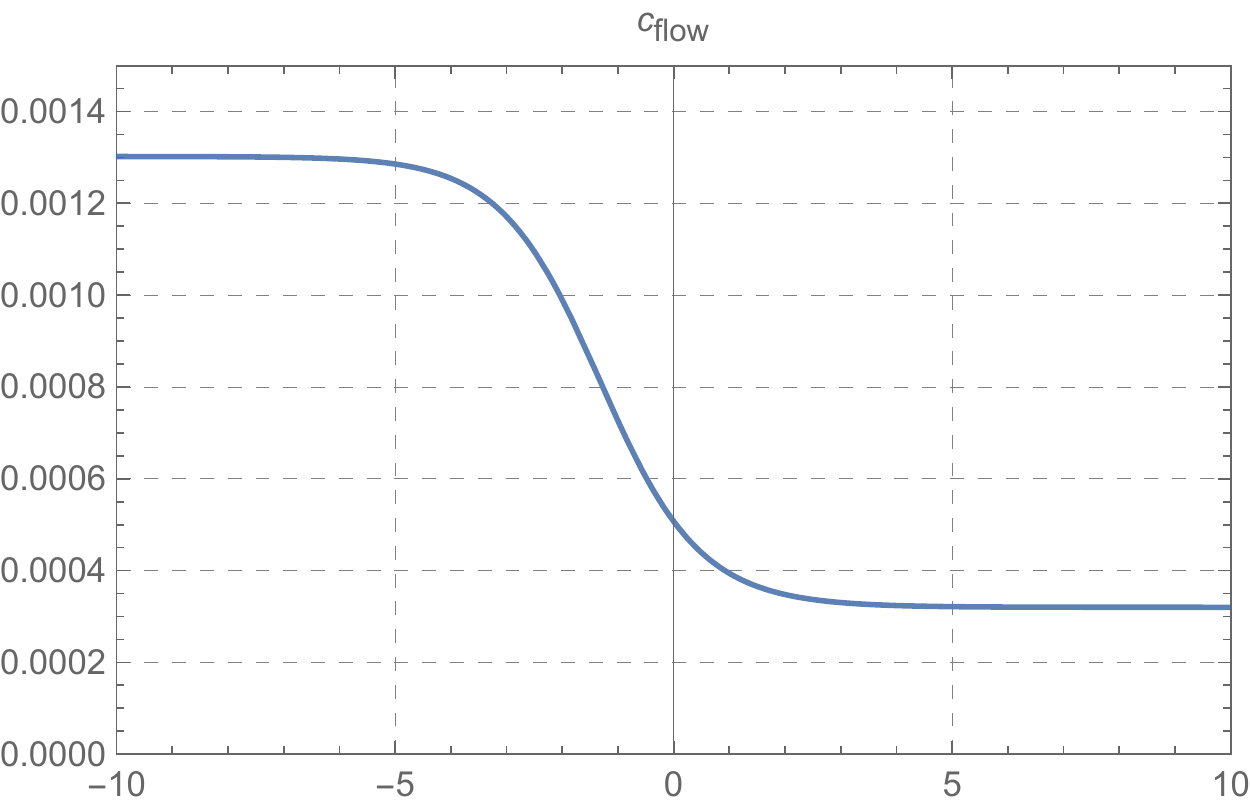}
        \caption{Flow central charge calculated for the numerical solutions. We clearly see the attainment of both IR and UV fixed points. }\label{centralflow}
    \end{figure}
This plot is very similar to the analogous quantity calculated from the Entanglement Entropy in reference \cite{GonzalezLezcano:2022mcd}.
The discussion about the factors in $\hat{\cal N}$ is similar to that in the previous subsection. The qualitative lessons for the QFT along the flow are similar. In summary, we have defined a quantity for an anisotropic QFT realising a flow across dimensions. It should be  nice  to tackle this same problem from a purely field theoretical perspective.

\subsection{A phenomenological proposal for the QFT}\label{pheno}

We wish to put forwards a proposal for the brane system and the field theory dual to the background in eqs.(\ref{metric-z})-(\ref{eqF0}). Our proposal is motivated by the 
scaling  of the holographic central charge, calculated in eqs.(\ref{chol4d}),(\ref{chol4dex1}),(\ref{chol4dex2}). We find that our proposal is at best provisional. In fact, whilst the scaling with free energy of the quiver parameters matches with the holographic computations,  the precise coefficients do not match. We suggest  possible modifications of our proposal to be studied in the future. 

We will be particularly interested on the fixed point solution of eq.(\ref{fixedpoint}), that gives a family of backgrounds with an AdS$_5$ factor preserving four Poincare supercharges.

\begin{figure}[h!]
    \centering
    \includegraphics[width=1.1\linewidth]{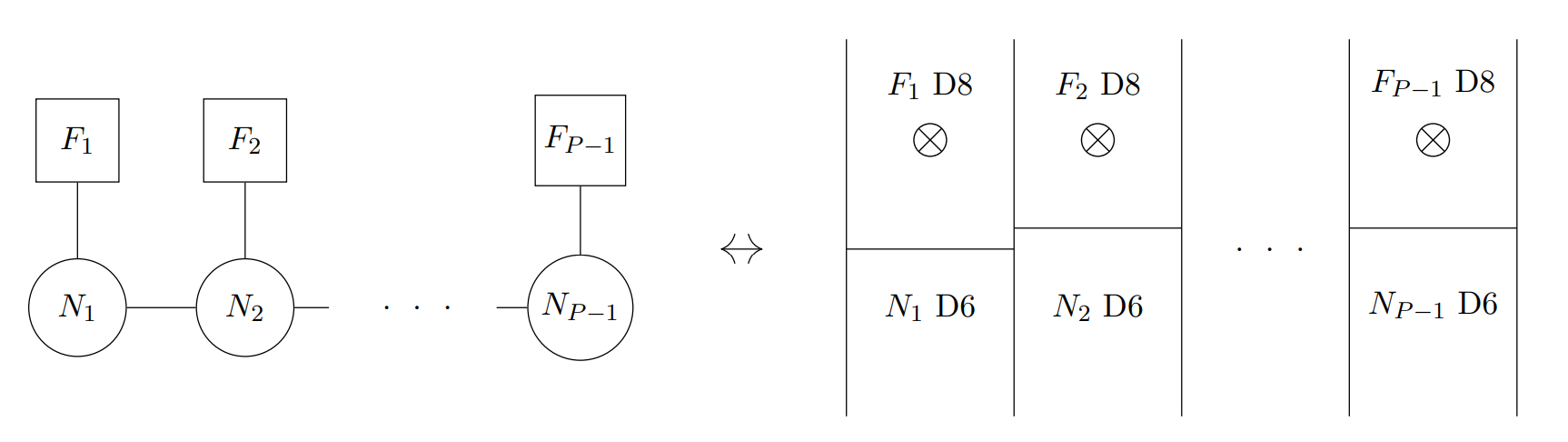}
    \caption{We show a {\it rough} approximation to the UV field theory. This would be just the torus reduction of a family of ${\cal N}=(1,0)$ six dimensional SCFTs. As we explain below, this picture is deformed by other brane stacks appearing.}
    \label{fig:xxbb}
\end{figure}

As we advanced in the previous section, for large values of the $r$-coordinate, the system is dual to a QFT that is to a good approximation six-dimensional.  In fact, using the solution of eq.(\ref{large-r-asymp}) in the background of eqs.(\ref{metric-z})-(\ref{eqF0}), we find that the space asymptotes to AdS$_7 \times S^2\times R_z$. The six dimensional SCFT is defined on a spacetime of the form $R^{1,3}\times H_2$. This forces the twisting of the SCFT (to partially preserve SUSY). The different fibrations in the Massive IIA background (encoded by the one form $\Theta_1$) are the effects of the twist in the holographic perspective. Following the formalism developed in the references \cite{sixd}-\cite{Nunez:2018ags}, we can {\it roughly} think that the field theory is encoded in a long linear quiver and Hanany-Witten set up of D6-D8-NS5 branes depicted in Figure \ref{fig:xxbb} 

The numbers $N_1, N_2,....,N_P$ and $F_1, ...., F_P$ must satisfy the relation $F_i = 2 N_i - N_{i-1}-N_{i+1}$ in order for the six dimensional field theory to be free of gauge anomalies. This condition is ensured by the function $\alpha''(z)$ chosen in eq.(\ref{alphasecond}). Indeed, note that from the fourth-derivative $\alpha^{(4)}(z)$ we derive the relation between flavours and colours required, see eq.(\ref{alpha4}). The function $\alpha(z)$, continuous and cubic by pieces is chosen such that the $z$-coordinate begins and ends smoothly. This is imposed by the conditions $\alpha(0)=\alpha(P)=0$. There is one UV conformal point for each choice of $\alpha(z)$. Of course, this is just  a {\it very rough} picture.

Indeed, these UV conformal points  are deformed (either by VEVs or by relevant operators). The dimension of these operators can be read from the near-AdS$_7$ expansion of the metric. These deformations (analogously, the presence of the fibrations) topologically twist the 6d CFT and trigger a RG flow, that ends in a CFT$_4$. As we discussed around eqs. (\ref{cargasns}), (\ref{chargesinterval}), a new set of NS and D6-branes appears due to the twisted compactification. These new branes lead to the presence of new gauge groups, not present in the original  linear quiver without twisting.

\begin{figure}[h!]
    \centering
    \includegraphics[width=1.1\linewidth]{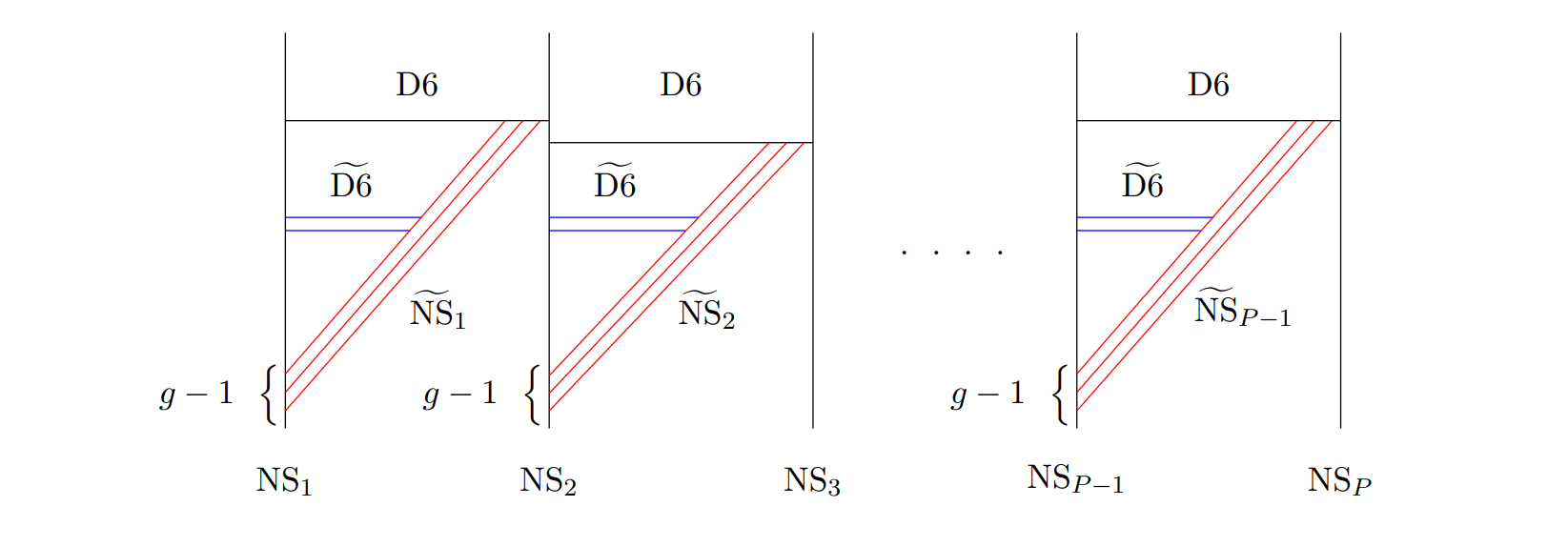}
    \caption{We draw the new set of branes--denoted with a tilde--which are induced by the twisting. These new branes give place to new gauge groups and new matter fields represented by strings extending between different stacks of branes.}
    \label{hananywittend6d8ns}
\end{figure}

Following this line of arguments,  our system is represented by the Hanany-Witten set up 
of Figure \ref{hananywittend6d8ns}.

As expressed in eq.(\ref{cargasns}), a new set of NS-five branes appears due to the twisting. These new $(g-1)\times P$ NS$_5$ branes
share the $R^{1,3}$ directions with the other branes and extend along the coordinates $(\theta_2,\phi_2)$. They are orthogonal to the original NS branes that extend along $(R^{1,3},\theta_1,\phi_1)$. This leads to the generation of new gauge groups. In fact, the  D6 branes originally extended along $(R^{1,3},\theta_1,\phi_1,z)$ can now extend  between the two stacks of NS-five branes. This generates a  new stack of D6 branes extended along $(R^{1,3},\theta_2,\phi_2,z)$. We associate with these the  new gauge groups. 
As calculated in eq.(\ref{chargesinterval}), there are $(g-1)\times N_l$ of them, for each interval. There are also D8 sources, playing the role of flavour groups. The system must cancel all gauge anomalies.  We emphasise  the $(g-1)\times P\times P$-replication of  the original quiver. This reflects in the free energy calculated in Section \ref{centralsection}. It should be interesting to carefully quantise the open strings between these different stacks of branes, to have a more concrete handle on the low energy QFT.

\begin{figure}[h!]
    \centering
    \includegraphics[width=0.9\linewidth]{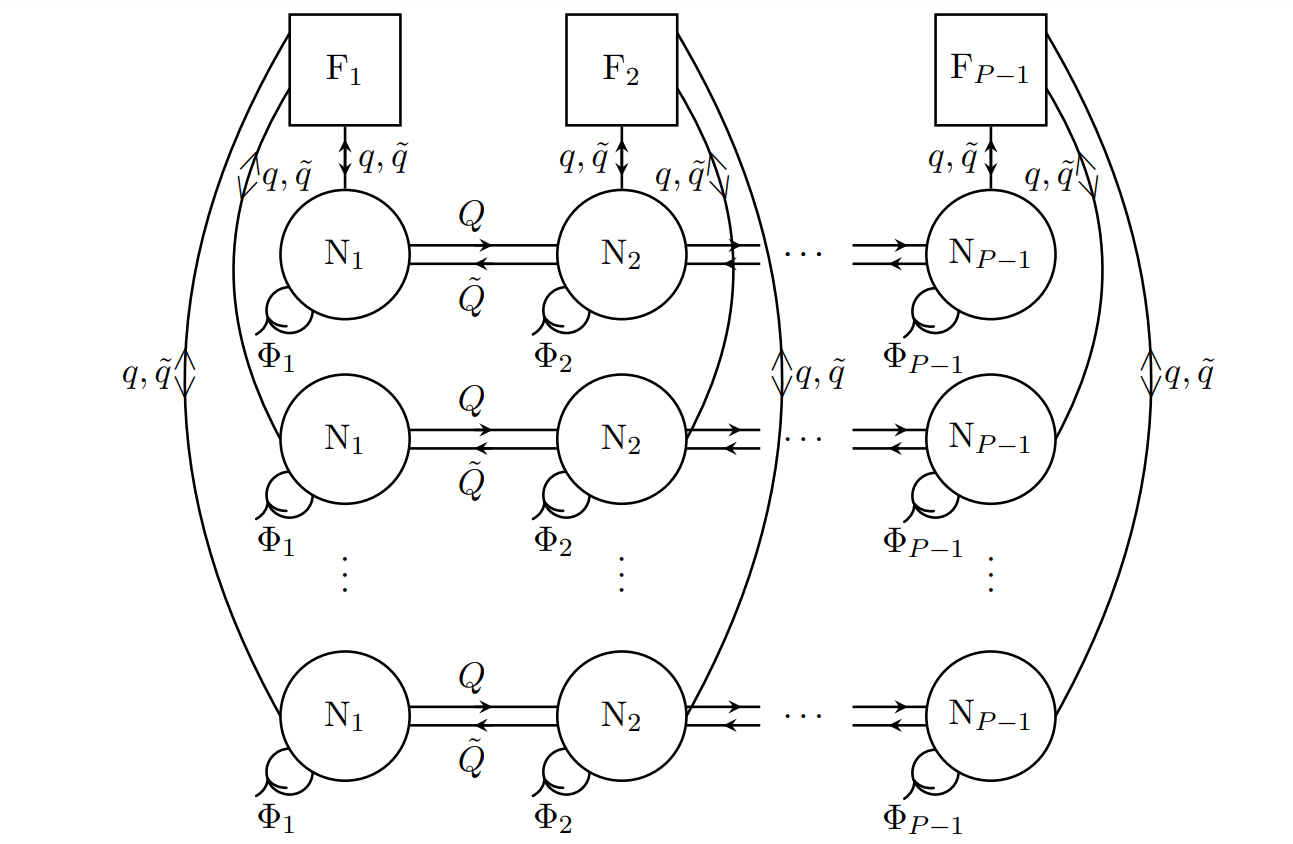}
        \caption{The proposed quiver field theory. The adjoint fields $\Phi_i$ get a mass. The bifundamentals fields are denoted by $(Q,\bar{Q})$, the fundamentals by $(q,\bar{q})$ and the vector multiplets represented by the circles. All the fields are in four dimensional ${\cal N}=1$ notation.}
    \label{quiver4d}
\end{figure}

A natural question that arises is the following:
can we find a Lagrangian description for the 4d QFT? If so, we could propose a (low energy) duality between a 4d Lagrangian QFT and a 6d SCFT on a geometry of the form $R^{1,3}\times \Sigma_{\kappa}$. This is perhaps too much to ask. We should be content  with a 4d quiver (even when strongly coupled) that captures aspects of the compactified 6d theory.
In fact,  the above analysis leads us to propose a quiver, that consist on $P^2(g-1)$ copies of the six-dimensional 'mother' theory. Of course, the system is now four dimensional and preserves ${\cal N}=1$ SUSY. In the usual notation (lines with arrows indicating chiral multiplets, circles denoting vector multiplets, boxes indicating flavour symmetries), we propose the quiver  in Figure \ref{quiver4d}. 


In Appendix \ref{sec-analysisqft}, we assign values to the anomalous dimensions and R-charges of each of these four dimensional fields.
Using those values we calculate beta functions and R-symmetry anomalies for each gauge group (finding vanishing values). Also a suitable superpotential is proposed, this indicates an interaction between different rows of the quiver in Figure \ref{quiver4d}. With the  R-charges assignation above, we also calculate $a_{CFT}$ and $c_{CFT}$, the $a,c$-central charges of the  quiver in Figure \ref{quiver4d}.  We compare some of  these results with the analogous  holographic ones in Section \ref{centralsection}. Whilst the scaling with the quiver parameters (number and rank of gauge nodes) is matched, the precise numerical coefficient does not coincide.

Let us close this section with three important comments.
\begin{itemize}
\item{It may be the case that other fields that do not contribute to the beta functions and the R-symmetry anomaly, but still contribute to the $a,c$ central charges, are present in our quiver of Figure \ref{quiver4d}. To decide if they should be there, we should compute the integral of the anomaly polynomial $I_8$ on the compactification manifold $\Sigma_{-}$. We get a hint at those fields, by comparing the material of Section \ref{centralsection} with that in Appendix \ref{sec-analysisqft}.}
\item{
The scaling with the number of nodes $P$ of the a-central charge (the free energy) is two powers higher than usual (see Appendix \ref{sec-analysisqft} for a field theoretical derivation of this result, and Section \ref{centralsection} for a holographic viewpoint on this). In fact, it actually scales as a six dimensional CFT.
The quiver we write is {\it deconstructing} two dimensions, in the sense of \cite{Arkani-Hamed:2001wsh}. What is interesting about this case is that this is occurring due to the twisting, whilst in \cite{Arkani-Hamed:2001wsh} it is an effect of going to a particular point in the moduli space, expanding, etc.}
\item{Finally, as explained in Appendix \ref{sec-analysisqft}, a mass term for $\Phi_i$ in eq.(\ref{superpot}) appears. This  generates a new set of $(P-1)-U(1)$'s. These are all anomalous, except for one. The R-symmetry we have taken is the one after this mixing has taken place. Importantly, this makes the beautiful result of Bobev and Crichigno \cite{Bobev:2017uzs}, not applicable here. In fact, the authors of \cite{Bobev:2017uzs} assume the absence of mixings. In this sense, the coefficient in their eq.(3.6) for the $a$ central charge, is not enforced on us\footnote{Special thanks to Nikolay Bobev for a discussion on this.}. }
\end{itemize}

Let us close this work by presenting some conclusions.

\section{Conclusions}\label{concl}
Let us start with a brief overview of the contents of this paper.  

In Section \ref{sectiongeometry}, we construct an infinite family of Massive Type IIA backgrounds. These describe holographically the flow between a family of 
six-dimensional ${\cal N}=(1,0)$ SCFTs and four dimensional ${\cal N}=1$ SCFTs. 
The charges of the brane system are discussed. New sets of branes appear induced by the twisted-compactification and are carefully discussed.

 In Section \ref{QFTsection}, we present and calculate two quantities that measure the number of degrees of freedom along the flow. One of these quantities coincides with the holographic central charge for the 4d SCFTs, and diverges in the UV.  The second quantity detects both fixed points,  it is monotonous,  its value being bigger in the IR than in the UV.   We also present a phenomenological proposal for a family of dual four dimensional SCFTs of the linear quiver type. The form and the parameters characterising the quiver are inherited from the six dimensional description. The 4d QFTs are proposed to reach a conformal point at low energies compared to the inverse size of the compactification manifold. These 4d SCFTs are holographically described by an infinite family of Massive Type IIA backgrounds with an AdS$_5$ factor.
We use the quiver field theory description to calculate field theory observables--see Appendix \ref{sec-analysisqft} for details. Interestingly our field theoretical calculation of the  $a$ and $c$ central charges in comparison with the holographic one,  suggests that other fields should intervene in the dynamics, aside from those in the proposed quiver.

Detailed appendices complement the presentation. We trust that the reader wishing to work on these topics should enjoy and profit from them.

This paper suggest various interesting topics to work on, as follow up of the material presented here. We make a small list below.
\begin{itemize}
\item{It should be interesting to further study the two-sphere compactifications. These do not lead to conformal 4d fixed points. It is important to de-singularise those solutions.}
\item{It is important to study the string quantisation in the Hanany-Witten set up that arises as a result of the twisted compactification. The massless modes would give firm clues about the four dimensional quiver field theory, for which we  proposed a particular type of quiver.}
\item{The presence of extra fields contributing to the central charges (as emphasised in Appendix \ref{sec-analysisqft}) is one of the most urgent topics suggested by this work. May be a treatment along the lines of \cite{Bah:2021iaa} proves effective. In fact, involving the anomaly polynomials $I_8$ and $I_6$ for the SCFT$_6$ and SCFT$_4$ may show the need to add flip fields.}
\item{More generally, bringing together the field theoretical techniques of \cite{examples2}-\cite{Bah:2021iaa} with the holographic results expressed in this paper or in \cite{Legramandi:2021aqv}, seems like a problem to study. }
\item{It should be interesting to attempt this kind of compactifications with systems in different dimensions. The systematic study of examples might indicate dimension-dependent characteristics.}
\item{It should be interesting to exploit the recently discovered plethora of embeddings of 5d supergravity into Massive IIA \cite{Couzens:2022aki}. Finding interesting compactifications of our family of backgrounds, black holes, and other solutions may illuminate various aspects of the ${\cal N}=1$, four dimensional  family of SCFTs.}
\end{itemize}
We hope to return to these problems soon.

 \section*{Acknowledgments: }
The contents and presentation of this work much benefitted from extensive discussion with various colleagues.  We would like to make a special mention to Nikolay Bobev, Stefano Cremonesi and Alessandro Tomasiello. We would also like to thank  very useful conversations with Mohammad Akhond, Andrea Legramandi, Yolanda Lozano, Daniel Thompson.

For the purpose of open access, the authors have applied a Creative Commons Attribution (CC BY) licence to any Author Accepted Manuscript version arising.

We are supported by  STFC  grant  ST/T000813/1.
The work of R.S. is supported by STFC grant ST/W507878/1.

\appendix
\section{Appendix 1}\label{appendixone}

In this appendix we outline in detail the construction of the infinite family of solutions of Massive Type IIA used in this paper. We begin by considering 7D $SU(2)$ Topologically Massive gauged Supergravity. We propose an ansatz and derive the BPS equations from the SUSY variations. Finally, we outline the uplift of the 7D background to 10D Massive Type IIA,  following the procedure used in \cite{Passias:2015gya}.

{\it Starting in Appendix \ref{backan} we change slightly the notation respect to the main body of the paper. We denote with a prime $()'$ the derivative with respect to the $r$-coordinate.}

\subsection{7D $SU(2)$ Topologically Massive Gauged Supergravity}

The Lagrangian, in string frame, of the bosonic fields of the 7D $SU(2)$ Topologically Massive Supergravity is given by,
    \begin{equation}\label{TopMassiveL}
    \begin{aligned}
        \mathcal{L} &= \sqrt{-g}e^{-2\Phi}\left[ R - \frac{1}{8} F^{i}_{\mu\nu}F^{\mu\nu\, i} + 4\partial^{\mu}\Phi\partial_{\mu}\Phi -\left( \frac{m^{2}}{2}e^{-4\Phi} - 4m\,e^{-2\Phi}-4 \right)\right]\\
        &~~\phantom{=} -\frac{1}{2}e^{2\Phi}\ast G_{4}\wedge G_{4} + \frac{1}{4}F^{i}\wedge F^{i} \wedge B_{3} - \frac{m}{2}G_{4}\wedge B_{3},
    \end{aligned}
    \end{equation}
    
where $G_{4}=dB_{3}$, $F = dA + i A\wedge A$, and $A$ is a $su(2)$ valued gauge 1-form, i.e. $A=A^{i} T^{i}$. The SUSY variations of the fermionic fields of the theory are
    \begin{align}
        \delta \lambda &= \left[ \Gamma^{\mu}\partial_{\mu}\Phi -\frac{i}{4}\Gamma^{\mu\nu}F_{\mu\nu} + \frac{1}{48} e^{2\Phi} \Gamma^{\mu\nu\lambda\rho}G_{\mu\nu\lambda\rho} - m\, e^{-2\Phi}+1 \right]\epsilon,\\
        \delta \Psi_{\mu} &= \left[ D_{\mu} + iA_{\mu} - \frac{i}{2}\Gamma^{\nu}F_{\mu\nu} + \frac{1}{96}e^{2\Phi}\Gamma^{\phantom{\mu}\nu\lambda\rho\sigma}_{\mu}G_{\nu\lambda\rho\sigma} - \frac{m}{4}e^{-2\Phi}\Gamma_{\mu} \right]\epsilon,
    \end{align}
    
where the covariant derivative is given by
    \begin{equation}
        D_{\mu}\epsilon = \left(\partial_{\mu} + \frac{1}{4}\omega_{\mu}^{\phantom{\mu}ab}\Gamma_{ab} \right)\epsilon.
    \end{equation}

Here $\mu,\nu$ are spacetime indices while $a,b$ are tangent space ones. 

We want to derive the BPS equations using a method similar to the one showed in \cite{Paredes:2004xw}, where the author studied the same Supergravity theory but for vanishing topological mass ($m=0$).  For this, we need to redefine our 1-form gauge field as follows: first we express the $su(2)$ generators, $T^{i}$, in terms of the Pauli matrices, $\sigma^{i}$, as $T^{i}=\frac{\sigma^{i}}{2}$. Then we derive the expression for the components of the field strength along the algebra generators. Using $A\wedge A = \frac{1}{2} [A,A]$, we have
    \begin{equation}
        F = dA + \frac{i}{2} [A,A] = dA^{i} \frac{\sigma^{i}}{2} + \frac{i}{8} \left[\sigma^{j},\sigma^{k}\right]^{i} A^{j}\wedge A^{k}.
    \end{equation}
    
Using the algebra of the Pauli matrices, $\left[\sigma^{i},\sigma^{j}\right]=2i\,\varepsilon^{ijk} \sigma^{k}$, and expanding $F = F^{i}\frac{\sigma^{i}}{2}$, we get
    \begin{equation}
        F^{i} = dA^{i} - \frac{1}{2}\varepsilon^{ijk} A^{j}\wedge A^{k}.
    \end{equation} 
    
In order to get both the same field strength and SUSY variations as in \cite{Paredes:2004xw} in the limit $m=0$, we redefine $A^{i}\rightarrow -A^{i}$, which leads to
    \begin{equation}
        F^{i} = - \left(dA^{i} + \frac{1}{2}\varepsilon^{ijk} A^{j}\wedge A^{k}  \right) = -F'^{\,i}.
    \end{equation}
    
Note that this change doesn't affect the Lagrangian since it is quadratic in F, while the SUSY variations now change, since they are linear in $A$ and $F$. Using these redefinitions and expanding in the algebra generators, one finds
    \begin{align}
        \delta \lambda &= \left[ \Gamma^{\mu}\partial_{\mu}\Phi +\frac{i}{8}\Gamma^{\mu\nu}F^{i}_{\mu\nu}\sigma^{i} + \frac{1}{48} e^{2\Phi} \Gamma^{\mu\nu\lambda\rho}G_{\mu\nu\lambda\rho} - m\, e^{-2\Phi}+1 \right]\epsilon,\\
        \delta \Psi_{\mu} &= \left[ D_{\mu} - \frac{i}{2}A^{i}_{\mu}\sigma^{i} + \frac{i}{4}\Gamma^{\nu}F^{i}_{\mu\nu}\sigma^{i} + \frac{1}{96}e^{2\Phi}\Gamma^{\phantom{\mu}\nu\lambda\rho\sigma}_{\mu}G_{\nu\lambda\rho\sigma} - \frac{m}{4}e^{-2\Phi}\Gamma_{\mu} \right]\epsilon,
    \end{align}
    
note that we dropped the $'$ of $F'$.

\subsubsection{Background Ansatz}\label{backan}

For the background metric we consider the following fibred version of $\mathbb{R}^{3,1}\times\Sigma_{k}\times \mathbb{R}$
    \begin{equation}\label{space1}
        ds^{2} = e^{2f(r)} dx^{2}_{3,1} + e^{2h(r)}\left(d\theta^{2} 
        + \frac{1}{k}\sin^{2}(\sqrt{k}\theta)d\phi^{2}\right) + dr^{2},
    \end{equation}
    
where $k$ is proportional to the curvature of $\Sigma_{k}$. Taking $k=-1,0,1$, gives $\Sigma_{k}=H^{2},\mathbb{R}^{2}$ or $S^{2}$, respectively. Because we are interested in studying twisted compactifications we will only consider the cases $k=\pm 1$.

The rest of the background fields are $B_{3}=0$, $\Phi=\Phi(r)$ and
    \begin{equation}
        A^{1} = - k\,a(r)d\theta, \quad
        A^{2} = \frac{1}{\sqrt{k}}a(r)\sin(\sqrt{k}\theta)d\phi, \quad
        A^{3} = -\frac{1}{k}\cos(\sqrt{k}\theta) d\phi,
    \end{equation} 
    
which leads to the field strengths\footnote{In the appendixes, we have denoted the derivative with respect to $r$ with a prime. In the main body of the paper we have reserved the primes for derivatives with respect to $z$, whilst denoting the $r$-derivatives with a dot.}
    \begin{equation}
        F^{1} = -k\,a'(r)\, dr\wedge d\theta, \quad
        F^{2} = \frac{1}{\sqrt{k}}a'(r)\sin(\sqrt{k}\theta)\, dr\wedge d\phi,\quad
        F^{3} = \Big(1-k\,a^{2}(r)\Big)\frac{\sin(\sqrt{k}\theta)}{\sqrt{k}}\, d\theta\wedge d\phi.
    \end{equation}   

In what follows we will need the vielbeins
    \begin{equation}
        e^{\hat{m}} = e^{f(r)} dx^{m} , \quad
        e^{\hat{\theta}} = e^{h(r)} d\theta , \quad
        e^{\hat{\phi}} =\frac{1}{\sqrt{k}} e^{h(r)}\sin(\theta) d\phi , \quad
        e^{\hat{r}} = dr,
    \end{equation}  
and the spin connection
    \begin{equation}
        \omega^{\hat{m}\hat{r}} = f'(r) \,e^{f(r)} dx^{m} , \quad
        \omega^{\hat{\theta}\hat{\phi}} = -\cos(\sqrt{k}\theta) d\phi, \quad 
        \omega^{\hat{\theta}\hat{r}} = h'(r)\, e^{h(r)} d\theta, \quad
        \omega^{\hat{\phi}\hat{r}} =\frac{1}{\sqrt{k}} h'(r)\, e^{h(r)} \sin(\sqrt{k}\theta) d\phi.
    \end{equation}
In the vielbein basis, we have
    \begin{equation}
        A^{1} = - k\,a(r)\,e^{-h(r)} e^{\hat{\theta}}, \quad
        A^{2} = a(r)\,e^{-h(r)}e^{\hat{\phi}}, \quad
        A^{3} = \frac{1}{\sqrt{k}}e^{-h(r)} \text{cotg}(\sqrt{k}\theta) e^{\hat{\phi}},
    \end{equation}
    \begin{equation}
        F^{1} = k\, a'(r) e^{-h(r)} e^{\hat{\theta}}\wedge e^{\hat{r}}, \quad
        F^{2} = - a'(r) e^{-h(r)} e^{\hat{\phi}}\wedge e^{\hat{r}}, \quad
        F^{3} = \Big(1-k\,a^{2}(r)\Big) e^{-2h(r)} e^{\hat{\theta}}\wedge e^{\hat{\phi}},
    \end{equation}
    \begin{equation}
        \omega^{\hat{m}\hat{r}} = f'(r) e^{\hat{m}} , \quad
        \omega^{\hat{\theta}\hat{\phi}} = -\sqrt{k}\, e^{-h(r)} \text{cotg}(\sqrt{k}\theta) e^{\hat{\phi}} , \quad
        \omega^{\hat{\theta}\hat{r}} = h'(r) e^{\hat{\theta}} ,\quad
        \omega^{\hat{\phi}\hat{r}} = h'(r) e^{\hat{\phi}}.
    \end{equation}
    
\subsection{From SUSY Variations to BPS Equations}

Substituting the above ansatz into the expressions for the SUSY variations, one derives BPS equations as outlined below. We begin with the Dilatino variation and move onto the Gravitino variations.

\subsubsection{Dilatino Variation}

Substituting the ansatz into the Dilatino variation leads to
    \begin{equation}
        \delta\lambda = \left[ \Gamma_{\hat{r}} \Phi'(r) + \frac{i}{4}\left(
        k\, a'(r) e^{-h(r)} \Gamma_{\htheta\hr}\sigma^{1} -a'(r) e^{-h(r)} \Gamma_{\hphi \hr} \sigma^{2} + \Big(1-k\,a^{2}(r)\Big)e^{-2h(r)}\Gamma_{\htheta\hphi}\sigma^{3} \right)-m\,e^{-2\Phi(r)} +1 \right]\epsilon.
    \end{equation}
    
We now impose the projection
    \begin{equation}\label{projection}
        \Gamma_{\htheta\hphi} \epsilon = \frac{1}{k}\sigma^{1}\sigma^{2}\epsilon.
    \end{equation}

Recalling that the Pauli matrices satisfy $\sigma^{1}\sigma^{2}\sigma^{3}=i$, we have    
    \begin{equation}
        \delta\lambda = \left[ \Gamma_{\hat{r}} \Phi'(r) + \frac{i}{4}a'(r) e^{-h(r)}\left(
        k\, \Gamma_{\htheta\hr}\sigma^{1} - \Gamma_{\hphi \hr} \sigma^{2}\right) -\frac{1}{4k}\Big(1-k\,a^{2}(r)\Big)e^{-2h(r)} -m\,e^{-2\Phi(r)} +1 \right]\epsilon,
    \end{equation}
    
then, by multiplying by $\Gamma_{\hr}$, we get
    \begin{equation}
        \Gamma_{\hr}\delta\lambda = \left[ \Phi'(r) - \frac{i}{4}a'(r) e^{-h(r)}\left(
        k\, \Gamma_{\htheta}\sigma^{1} - \Gamma_{\hphi} \sigma^{2}\right) +\left(1-m\,e^{-2\Phi(r)} -\frac{1}{4k}\Big(1-k\,a^{2}(r)\Big)e^{-2h(r)}\right)\Gamma_{\hr} \right]\epsilon.
    \end{equation}

Using the projection \eqref{projection} and the fact that $k=\pm 1$ one can show that
    \begin{equation}\label{relation1}
        \left(k\, \Gamma_{\htheta}\sigma^{1} - \Gamma_{\hphi} \sigma^{2}\right)\epsilon 
        = 2k \Gamma_{\htheta}\sigma^{1}\epsilon.
    \end{equation}
    
Replacing this in the variation, yields    
    \begin{equation}\label{varDil}
        \Gamma_{\hr}\delta\lambda = \left[ \Phi'(r) - \frac{i}{2}k\,a'(r) e^{-h(r)}\Gamma_{\htheta}\sigma^{1} +\left(1-m\,e^{-2\Phi(r)} -\frac{1}{4k}\Big(1-k\,a^{2}(r)\Big)e^{-2h(r)}\right)\Gamma_{\hr} \right]\epsilon.
    \end{equation}

\subsubsection{Gravitino Variation}

We now switch our focus to the Gravitino variations, focusing on the $\hm$, $\htheta$, $\hphi$ and $\hr$ components in turn. From the $m=0,1,2,3$ components of the variation of the gravitino in tangent space indices, we have
    \begin{equation}
        \delta\psi_{\hm} = \left( \frac{1}{2} f'(r)\Gamma_{\hm\hr} -\frac{m}{4}e^{-2\Phi(r)}\Gamma_{\hm} \right)\epsilon,
    \end{equation}
and multiplying by $2\Gamma_{\hr\hm}$, gives
    \begin{equation}\label{varGravm}
        2\Gamma_{\hr\hm}\delta\psi_{\hm} = \left( f'(r) -\frac{m}{2}e^{-2\Phi(r)}\Gamma_{\hr} \right)\epsilon. 
    \end{equation}   
     
From the $\htheta$ component, we have
    \begin{equation}
        \delta\psi_{\htheta} = \left[\frac{1}{2}h'(r) \Gamma_{\htheta\hr} +\frac{i}{2}k\,a(r)\,e^{-h(r)}\sigma^{1} +\frac{i}{4}\left(k\,a'(r)e^{-h(r)}\Gamma_{\hr} \sigma^{1} + \Big(1-k\,a^{2}(r)\Big)e^{-2h(r)}\Gamma_{\hphi}\sigma^{3} \right)-\frac{m}{4}e^{-2\Phi(r)}\Gamma_{\htheta}\right]\epsilon.
    \end{equation}    
    
Multiplying by $2\Gamma_{\hr\htheta}$, we get     
    \begin{equation}
    \hspace{-6.5mm}
        2\Gamma_{\hr\htheta}\delta\psi_{\htheta} = \left[h'(r)  +i\,k\,a(r)\,e^{-h(r)}\Gamma_{\hr\htheta}\sigma^{1} +\frac{i}{4}\left(-k\,a'(r)e^{-h(r)}\Gamma_{\htheta} \sigma^{1} + \Big(1-k\,a^{2}(r)\Big)e^{-2h(r)}\Gamma_{\hr}\Gamma_{\htheta\hphi}\sigma^{3} \right)-\frac{m}{4}e^{-2\Phi(r)}\Gamma_{\hr}\right]\epsilon.
    \end{equation} 
    
Using the projection \eqref{projection} leads to    
    \begin{equation}\label{varGravtheta}
        2\Gamma_{\hr\htheta}\delta\psi_{\htheta} = \left[h'(r)  +i\,k\,a(r)\,e^{-h(r)}\Gamma_{\hr\htheta}\sigma^{1} -\frac{i}{2}k\,a'(r)e^{-h(r)}\Gamma_{\htheta} \sigma^{1}-\frac{1}{2k} \Big(1-k\,a^{2}(r)\Big)e^{-2h(r)}\Gamma_{\hr}-\frac{m}{2}e^{-2\Phi(r)}\Gamma_{\hr}\right]\epsilon.
    \end{equation}    
      
From the $\hphi$ component we have
    \begin{equation}
    \begin{aligned}
        \delta\psi_{\hphi} = &\left[ \frac{1}{2}\left( h'(r) \Gamma_{\hphi\hr} - \sqrt{k}e^{-h(r)}\cotg(\sqrt{k}\theta) \Gamma_{\htheta\hphi} \right) -\frac{i}{2}\left( a(r)\,e^{-h(r)}\sigma^{2} +\frac{1}{\sqrt{k}}e^{-h(r)}\cotg(\sqrt{k}\theta)\sigma^{3}\right) \right.\\
        &~~\left.+\frac{i}{4}\left( -a'(r)e^{-h(r)}\Gamma_{\hr}\sigma^{2} -\Big(1-k\,a^{2}(r)\Big)e^{-2h(r)}\Gamma_{\htheta}\sigma^{3} \right) - \frac{m}{4}e^{-2\Phi(r)}\Gamma_{\hphi}\right]\epsilon.
    \end{aligned}
    \end{equation}
    
Using \eqref{projection}, the second and the fourth term cancel. This is the effect of the topological twist. We are left with
    \begin{equation}
        \delta\psi_{\hphi} = \left[ \frac{1}{2} h'(r) \Gamma_{\hphi\hr}  -\frac{i}{2} a(r)\,e^{-h(r)}\sigma^{2} +\frac{i}{4}\left( -a'(r)e^{-h(r)}\Gamma_{\hr}\sigma^{2} -\Big(1-k\,a^{2}(r)\Big)e^{-2h(r)}\Gamma_{\htheta}\sigma^{3} \right) - \frac{m}{4}e^{-2\Phi(r)}\Gamma_{\hphi}\right]\epsilon,
    \end{equation}
    
then multiplying by $2\Gamma_{\hr\hphi}$, gives
    \begin{equation}
        2\Gamma_{\hr\hphi}\delta\psi_{\hphi} = \left[ h'(r)  -i\, a(r)\,e^{-h(r)}\Gamma_{\hr\hphi}\sigma^{2} +\frac{i}{2}\left( a'(r)e^{-h(r)}\Gamma_{\hphi}\sigma^{2} +\Big(1-k\,a^{2}(r)\Big)e^{-2h(r)}\Gamma_{\hr}\Gamma_{\htheta\hphi}\sigma^{3} \right) - \frac{m}{2}e^{-2\Phi(r)}\Gamma_{\hr}\right]\epsilon.
    \end{equation}
    
Using the projection again, leads to
    \begin{equation}
        2\Gamma_{\hr\hphi}\delta\psi_{\hphi} = \left[ h'(r)  -i\, a(r)\,e^{-h(r)}\Gamma_{\hr\hphi}\sigma^{2} +\frac{i}{2} a'(r)e^{-h(r)}\Gamma_{\hphi}\sigma^{2} -\frac{1}{2k}\Big(1-k\,a^{2}(r)\Big)e^{-2h(r)} - \frac{m}{2}e^{-2\Phi(r)}\Gamma_{\hr}\right]\epsilon,
    \end{equation}   
    
and due to \eqref{relation1}, we have $k \Gamma_{\htheta}\sigma^{1}\epsilon=-\Gamma_{\hphi}\sigma^{2}\epsilon$, giving
    \begin{equation}
        2\Gamma_{\hr\hphi}\delta\psi_{\hphi} = \left[ h'(r) +  i\,k\, a(r)\,e^{-h(r)}\Gamma_{\hr\htheta}\sigma^{1} -\frac{i}{2}k\, a'(r)e^{-h(r)}\Gamma_{\htheta}\sigma^{1} -\frac{1}{2}\Big(1-k\,a^{2}(r)\Big)e^{-2h(r)} - \frac{m}{2}e^{-2\Phi(r)}\Gamma_{\hr}\right]\epsilon,
    \end{equation} 
    
 so we see that $\Gamma_{\hr\htheta}\delta\psi_{\htheta} = \Gamma_{\hr\hphi}\delta\psi_{\hphi}$. \\
 
 Finally, from the $\hr$ component of the gravitino variation, we have
    \begin{equation}
        \delta\psi_{\hr}  = \left[\partial_{r} +\frac{i}{4}\left( -k\,a'(r) e^{-h(r)}\Gamma_{\htheta}\sigma^{1} + a'(r)e^{-h(r)} \Gamma_{\hphi}\sigma^{2} \right) 
        - \frac{m}{4}e^{-2\Phi(r)}\Gamma_{\hr}\right]\epsilon,
    \end{equation}
    
which can be rewritten as
    \begin{equation}
        \delta\psi_{\hr}  = \left[\partial_{r} +\frac{i}{4}a'(r)e^{-h(r)}\left( -k \,\Gamma_{\htheta}\sigma^{1} + \Gamma_{\hphi}\sigma^{2} \right) 
        - \frac{m}{4}e^{-2\Phi(r)}\Gamma_{\hr}\right]\epsilon,
    \end{equation}
    
and using the projection, we find
    \begin{equation}
         \delta\psi_{\hr}  = \left[\partial_{r} -\frac{i}{2}k\,a'(r)e^{-h(r)} \Gamma_{\htheta}\sigma^{1}  - \frac{m}{4}e^{-2\Phi(r)}\Gamma_{\hr}\right]\epsilon.
    \end{equation}
    
After imposing $\delta\lambda=0$ and $\delta\psi_{\mu}=0$, the results \eqref{varDil},\eqref{varGravm} and \eqref{varGravtheta} lead to the following BPS equations
    \begin{align}
        &\Phi'(r)\epsilon - \frac{i}{2}k\,a'(r) e^{-h(r)}\Gamma_{\htheta}\sigma^{1}\epsilon +\left(1-m\,e^{-2\Phi(r)} -\frac{1}{4k}\Big(1-k\,a^{2}(r)\Big)e^{-2h(r)}\right)\Gamma_{\hr}\epsilon = 0, \label{BPS1}\\
        &f'(r)\epsilon -\frac{m}{2}e^{-2\Phi(r)}\Gamma_{\hr}\epsilon = 0,\label{BPS2}\\
        & h'(r)\epsilon  +i\,k\,a(r)\,e^{-h(r)}\Gamma_{\hr\htheta}\sigma^{1}\epsilon -\frac{i}{2}k\,a'(r)e^{-h(r)}\Gamma_{\htheta} \sigma^{1}\epsilon
        -\frac{1}{2k} \Big(1-k\,a^{2}(r)\Big)e^{-2h(r)}\Gamma_{\hr}\epsilon
        -\frac{m}{2}e^{-2\Phi(r)}\Gamma_{\hr}\epsilon = 0,\label{BPS3}\\
        &\partial_{r}\epsilon = \frac{i}{2}k\,a'(r)e^{-h(r)}\Gamma_{\htheta}\sigma^{1}\epsilon + \frac{m}{4}e^{-2\Phi(r)}\Gamma_{\hr}\epsilon.
    \end{align}

\subsubsection{Rearranging the SUSY variations}

Following the procedure given in \cite{Paredes:2004xw}, we now rearrange the SUSY variations to obtain the BPS equations. For this, we use the properties of the $\Gamma$ matrices to simplify the equations obtained in the previous subsection.

Let us start by rewriting the dilatino variation \eqref{varDil} as
    \begin{equation}\label{BPS1beta}
        \Gamma_{\hr} \epsilon = \beta\epsilon + \tilde{\beta} i\Gamma_{\htheta}\sigma^{1}\epsilon,
    \end{equation}
    
with
    \begin{align}
        \beta &= \frac{-\Phi'(r)}{1-m\,e^{-2\Phi(r)} -\frac{1}{4k}(1-k\,a^{2}(r))e^{-2h(r)}},\label{beta}\\
        \tilde{\beta} &= \frac{\frac{1}{2}k\,a'(r)e^{-h(r)}}{1-m\,e^{-2\Phi(r)} -\frac{1}{4k}(1-k\,a^{2}(r))e^{-2h(r)}}. \label{betatilde}
    \end{align}
    
By applying $\Gamma_{\hr}$ to \eqref{BPS1beta}, we get
    \begin{equation}
        \epsilon = \beta\, \Gamma_{\hr}\epsilon - i\tilde{\beta}\,\Gamma_{\htheta}\sigma^{1}\Gamma_{hr}\epsilon.
    \end{equation}
    
Using \eqref{BPS1beta}, gives    
    \begin{equation}
        \epsilon = \beta \left(  \beta\,\epsilon + \tilde{\beta}\, i\Gamma_{\htheta}\sigma^{1}\epsilon \right) - i\tilde{\beta}\,\Gamma_{\htheta}\sigma^{1}\left(  \beta\,\epsilon + \tilde{\beta}\, i\Gamma_{\htheta}\sigma^{1}\epsilon \right),
    \end{equation}
    
and by simplifying, we obtain
    \begin{equation}
        \epsilon = (\beta^{2}+\tilde{\beta}^{2})\epsilon,
    \end{equation}
    
from which we see that
    \begin{equation}
        \beta^{2}+\tilde{\beta}^{2} = 1.
    \end{equation}
    
Multiplying \eqref{BPS2} by $\Gamma_{\hr}$ we get
    \begin{equation}\label{VarGm2}
        f'(r) \Gamma_{\hat{r}}\epsilon = \frac{m}{2}\,e^{-2\Phi(r)}\epsilon.
    \end{equation}

Combining this with \eqref{BPS2} leads to
    \begin{equation}
        \left(f'^{\,2}(r) - \frac{m^{2}}{4}e^{-4\Phi(r)}\right)\epsilon=0,
    \end{equation} 
    
which means that
    \begin{equation}\label{df1}
        f'(r) = \pm \frac{m}{2}\,e^{-2\Phi(r)}.
    \end{equation}   

Now we go back to \eqref{VarGm2} and substitute in \eqref{BPS1beta} as follows
    \begin{equation}
        f'(r) \left(\beta + \tilde{\beta}\,i\Gamma_{\hat{\theta}}\sigma^{1}\right)\epsilon - \frac{m}{2}\, e^{-2\Phi(r)}=0.
    \end{equation}
    
Collecting terms with and without $i\Gamma_{\hat{\theta}}\sigma^{1}$, we get
    \begin{equation}
        \left(f'(r)\,\beta - \frac{m}{2}\,e^{-2\Phi(r)}\right)\epsilon + f'(r)\, \tilde{\beta}\,i\Gamma_{\hat{\theta}}\sigma^{1} = 0,
    \end{equation}
    
from where we get two equations
    \begin{align}
        f'(r)\beta &=  \frac{m}{2}\,e^{-2\Phi(r)},\label{df2}\\
        f'(r)\tilde{\beta} &=0. \label{betat}
    \end{align}
    
From \eqref{df1} and \eqref{df2} we get $\beta=\pm 1$, and from \eqref{betat} we get $\tilde{\beta} = 0$, which in \eqref{betatilde} gives 
    \begin{equation}
        a'(r)=0,
    \end{equation}
    
and hence, \eqref{BPS1beta} becomes
    \begin{equation}\label{projection2}
        \Gamma_{\hat{r}}\epsilon = \beta \epsilon.
    \end{equation}

Now we turn to \eqref{BPS3}. By replacing \eqref{projection2} and $a'(r)=0$, we get
    \begin{equation}
         h'(r)\epsilon - i\beta\,k\,a(r)\,e^{-h(r)}\Gamma_{\htheta}\sigma^{1}\epsilon 
        -\beta \frac{1}{2k} \Big(1-k\,a^{2}(r)\Big)e^{-2h(r)}\epsilon
        -\beta\frac{m}{2}e^{-2\Phi(r)}\epsilon = 0.
    \end{equation}
    
Again, by collecting terms with and without $i\Gamma_{\hat{\theta}}\sigma^{1}$, we get
    \begin{equation}
        \left(h'(r)  -\beta \frac{1}{2k} \Big(1-k\,a^{2}(r)\Big)e^{-2h(r)} -\beta\frac{m}{2}e^{-2\Phi(r)}\right)\epsilon -i\beta\,k\,a(r)\,e^{-h(r)}\Gamma_{\hat{\theta}}\sigma^{1}\epsilon =0.
    \end{equation}
    
From which we get
    \begin{align}
        &h'(r)  -\beta \frac{1}{2k} \Big(1-k\,a^{2}(r)\Big)e^{-2h(r)} -\beta\frac{m}{2}e^{-2\Phi(r)} = 0,\label{eqdh}\\
        &a(r)\,e^{-h(r)} =0.
    \end{align}
    
From the second result, we necessarily get $a(r)=0$. After implementing this condition into \eqref{beta} and \eqref{eqdh}, we finally derive the following BPS equations for $f(r)$ (which we got from \eqref{df1}), $h(r)$ and $\Phi(r)$
    \begin{align}
        f'(r) &= \pm \frac{m}{2}\,e^{-2\Phi(r)},\label{BPSfAP}\\
        h'(r) &= \pm \frac{1}{2}\left( \frac{1}{k}e^{-2h(r)}+m\,e^{-2\Phi(r)} \right),\label{BPShAP}\\
        \Phi'(r) &= \pm\left(-1+\frac{1}{4k}e^{-2h(r)} + m\,e^{-2\Phi(r)}\right).\label{BPSpAP}
    \end{align}

The minus sign can be absorbed by $r\rightarrow -r$.

\subsection{Uplift to 10D Massive Type IIA}

\subsubsection{Einstein Frame and Normalizations}

The procedure which we will follow to uplift our 7D topologically massive solution is given in \cite{Passias:2015gya}, however,  the action we used for the 7D $\mathcal{N}=2$ $SU(2)$ Gauged SUGRA is written in string frame, while in \cite{Passias:2015gya} is written in Einstein frame, hence we move our action to Einstein frame and then we compare normalizations and relevant constants. To move to Einstein frame we use the transformation
    \begin{equation}
        g^{(E)}_{\mu\nu} = e^{-\alpha\Phi(r)}g^{(S)}_{\mu\nu},
    \end{equation}

in the action \eqref{TopMassiveL}. Then, by comparing with \cite{Passias:2015gya} we see that $\alpha=\frac{4}{5}$, $m=1$, $g=\sqrt{2}$, $h=\frac{1}{2}$ and that our Dilaton $\Phi(r)$ is related to the one in \cite{Passias:2015gya}, which we call $\varphi(r)$, by 
    \begin{equation}
        \varphi(r) = \frac{2\sqrt{10}}{5}\Phi(r).
    \end{equation}

Also, from the kinetic term for the gauge field, we see that in \cite{Passias:2015gya} the gauge field is normalised in a way in which the coupling constant appears in the definition of $F$ rather than as a coefficient of the kinetic term $F^{2}$, while in \eqref{TopMassiveL} we consider the opposite. To match conventions we re-scale our gauge field and field strength as
    \begin{equation} 
        A \rightarrow -\frac{1}{g}A, \quad F \rightarrow -\frac{1}{g}F.
    \end{equation}
    
With this, our 7D ansatz now reads
    \begin{equation}\label{ansatzG}
        ds^{2} = e^{-\frac{4}{5}\Phi(r)}\left( e^{2f(r)} ds^{2}_{3,1} + e^{2h(r)}\left(d\theta^{2} + \frac{1}{k}\sin^{2}(\sqrt{k}\theta)d\phi^{2}\right) + dr^{2} \right),
    \end{equation}
    
together with the gauge field
    \begin{equation}
        A_{1} = 0, \quad A_{2} = 0, \quad  A_{3} = \frac{1}{g}\frac{1}{k}\cos(\sqrt{k}\theta) d\phi,
    \end{equation}

which leads to
    \begin{equation}\label{ansatzF}
        F_{1} = 0, \quad  F_{2} = 0, \quad F_{3} = -\frac{1}{g}\Vol(\Sigma^{2}_{k}).
    \end{equation}

\subsubsection{Uplift of 7D Topologically Massive to Massive IIA}

The massive Type IIA ansatz is given by
    \begin{equation}
        ds^{2} = \frac{1}{\ell} X(r)^{-\frac{1}{2}} e^{2A(\rho)}ds^{2}_{7} + X(r)^{\frac{5}{2}} ds^{2}_{M_{3}},
    \end{equation}
    
with $\ell = 4$, $ds^{2}_{7}$ a solution of the 7D  $\mathcal{N}=2$ Gauged Supergravity presented in the previous section, and
    \begin{equation}
        ds^{2}_{M_{3}} = d\rho^{2} + \frac{1-x(\rho)^{2}}{16\omega}e^{2A(\rho)}Ds^{2}_{S^{2}},
    \end{equation}
    
where $X(r)$ is given by the 7D dilaton
    \begin{equation}
        X(r) = e^{\frac{2}{5}\Phi(r)},
    \end{equation}
    
and
    \begin{equation}
        \omega = X(r)^{5}(1-x(\rho)^{2})+x(\rho)^{2}.
    \end{equation}

The covariantized metric $Ds^{2}_{S^{2}}$ on the sphere is constructed as follows. First we consider the normal unitary vector to the sphere 
    \begin{equation}
        y^{i} = \left( \cos(\phi_{2})\sin(\theta_{2}), \sin(\phi_{2})\sin(\theta_{2}), \cos(\theta_{2}) \right),
    \end{equation}
    
then, the covariant line element is given by
    \begin{equation}
        Ds^{2}_{S^{2}}=Dy^{i}Dy^{i}, \quad Dy^{i} = dy^{i} +  g\epsilon^{ijk}y^{k}A^{k},
    \end{equation}
    
where $A^{i}$ is the component along the $i$th Pauli matrix of the 7D gauge field. This spacetime is supported by the 10D Dilaton $\Psi$  
    \begin{equation}
        e^{2\Psi} =  \frac{X(r)^{\frac{5}{2}}}{\omega} e^{2\psi(\rho)},
    \end{equation}

and the background forms
    \begin{align}
        B_{2} &= \left[\frac{1}{16 }e^{2A(\rho)}\frac{x(\rho)\sqrt{1-x(\rho)^2}}{\omega}\Vol_2
        -\frac{1}{2}e^{A(\rho)}d\rho\wedge\left(a-\frac{1}{g}y^{i}A^{i}\right)\right],\\
        F_{2} &= -q\left(\Vol_{2}+g\,y^{i}F^{i}\right)
        +\frac{1}{16\omega}\,F_0e^{2A(\rho)}x(\rho)\sqrt{1-x(\rho)^2}\Vol_{2},\\
        F_{4} &= \frac{\ell}{4}\left[-g\frac{q}{16\omega}e^{2A(\rho)}x(\rho)\sqrt{1-x(\rho)^{2}}y^{i}F^{i}\wedge \Vol_{2} -g\frac{q}{4}e^{A(\rho)}d\rho \wedge \epsilon^{ijk}F^{i}y^{j}Dy^{k}\right] + G_{4}\, \text{terms},
    \end{align}
where Vol$_{2}$ is the volume element of the covariantised sphere, $a = \frac{1}{2}\cos(\theta_{2})d\phi_{2}$ and $F_{0}$ is the Ramond mass (which is constant). Note that we have not written explicitly the terms proportional to the 7D 4-form, since we will not use them.\vspace{0.5cm}

This configuration is a solution to the equations of motion
    \begin{align}
        &\frac{1}{4}R+\nabla^{2} \Psi -(\nabla \Psi)^{2}-\frac{1}{8}H_{3}^{2}=0,\\
        & d F_{p} + H_{3}\wedge *F_{p-2} = 0,\\
        & d(e^{-2\Psi}* H_{3}) -\left(F_{0} *F_{2}+F_{2}\wedge *F_{4}+F_{4}\wedge F_{4}\right)=0,\\
        & R_{MN}+2\nabla_{M}\nabla_{N} \Psi -\frac{1}{2}(H^{2}_{3})_{MN}-\frac{1}{4}e^{2\Psi}\sum_{p} (F^{2}_{p})_{MN} ,
    \end{align}
with $p=2,4,6,8,10$, and
    \begin{equation}
        (F^{2}_{p})_{MN} = \frac{1}{(p-1)!}F_{M}^{\phantom{M}N_{1}...N_{p-1}}F_{MN_{1}...N_{p-1}},
    \end{equation}
and similarly for $ (H^{2}_{3})_{MN}$, provided $\psi(\rho)$, $x(\rho)$ and $A(\rho)$ satisfy
    \begin{align}
        \frac{d}{d\rho} \psi(\rho)&=\frac{1}{4}\frac{e^{-A(\rho)}}{\sqrt{1-x(\rho)^{2}}}\left(12x(\rho)+\Big(2x(\rho)^{2}-5\Big)F_{0}\,e^{A(\rho)+\psi(\rho)}\right),\\
         \frac{d}{d\rho} x(\rho)&=-\frac{1}{2}e^{-A(\rho)}\sqrt{1-x(\rho)^{2}}\Big(4+x(\rho)\,F_{0}\,e^{A(\rho)+\psi(\rho)}\Big),\\
        \frac{d}{d\rho} A(\rho)&=\frac{1}{4}\frac{e^{-A(\rho)}}{\sqrt{1-x(\rho)^2}}\left(4x(\rho)-F_{0}\, e^{A(\rho)+\psi(\rho)}\right).
\end{align}

\subsubsection{Explicit Uplift of the Interpolating Background}

We now write explicitly the field configuration for the uplift of \eqref{ansatzG}-\eqref{ansatzF}. The spacetime metric reads (here we rename $(\theta,\phi)\rightarrow(\theta_{1},\phi_{1})$ with respect to the 7D solution)
    \begin{equation}
    \begin{aligned}
        ds^{2} &= \frac{1}{\ell}X(r)^{-\frac{1}{2}}e^{2A(\rho)}e^{-\frac{4\Phi(r)}{5}}\left[e^{2f(r)}dx^{2}_{3,1}
        + dr^{2} + e^{2h(r)}\left( d\theta^{2}_{1} + \frac{1}{k}\sin^{2}(\sqrt{k}\theta_{1})d\phi^{2}_{1}\right)\right]\\
        &~~~~ + X(r)^{5/2}\left[ d\rho^{2} + \frac{1-x(\rho)^2}{16\omega}e^{2A(\rho)} \left(d\theta^{2}_{2}+\sin^{2}(\theta_{2})
        \left( d\phi_{2} - \frac{1}{k}\cos(\sqrt{k}\theta_{1})d\phi_{1}\right)^{2}\right)\right],
    \end{aligned}
    \end{equation}
    
with $\ell=4$, while the background forms read
\begin{equation}
     \hspace{-6.5mm}
    \begin{aligned}
        &B_{2} = \left(\frac{1}{16 \omega}e^{2A(\rho)}x(\rho)\sqrt{1-x(\rho)^2}\sin(\theta_{2})d\theta_{2} -\frac{1}{4}e^{A(\rho)}\cos(\theta_{2})d\rho \right)\wedge\left( d\phi_{2} -\frac{1}{k}\cos(\sqrt{k}\theta_{1})d\phi_{1} \right),\\
        &F_{2} = \frac{1}{4} e^{A(\rho)-\psi(\rho)}\sqrt{1-x(\rho)^{2}}\left[ \cos(\theta_{2})\Vol(\Sigma_{k}) -\Vol(S^{2}_{c}) \right]
         + \frac{1}{16\omega}\,F_0e^{2A(\rho)}x(\rho)\sqrt{1-x(\rho)^2}\, \Vol(S^{2}_{c}),\\
        &F_{4} = \frac{e^{3A(\rho)-\psi(\rho)}}{64\omega} \cos(\theta_{2})\, x(\rho)\Big(1-x(\rho)^{2}\Big)\Vol(\Sigma_{k})\wedge \Vol(S^{2}) + \frac{e^{2A(\rho)-\psi(\rho)}}{16} \sin^{2}(\theta_{2}) \sqrt{1-x^{2}(\rho)}\, d\rho\wedge d\phi_{2}\wedge \Vol(\Sigma_{k}),
    \end{aligned}
    \end{equation}
    
where 
    \begin{equation}
        \Vol(S^{2}) = \sin(\theta_{2})d\theta_{2} \wedge d\phi_{2},
    \end{equation}
    
is the volume of the 2-sphere of the internal manifold, while
    \begin{equation}
        \Vol(S^{2}_{c}) = \sin(\theta_{2})d\theta_{2} \wedge\left( d\phi_{2}-\frac{1}{k}\cos(k\theta_{1})d\phi_{1}\right),
    \end{equation}
    
is the volume of the covariantised 2-sphere. Note that 
    \begin{equation}
        \Vol(\Sigma_{k})\wedge\Vol(S^{2}_{c}) =\Vol(\Sigma_{k})\wedge\Vol(S^{2}).
    \end{equation}
    
\subsubsection{Page Fluxes}

In order to get a quantised number of charges, we need to consider the Page fluxes, given by
    \begin{equation}
        \hat{F}_{p} = F_{p}\wedge e^{-B_{2}},
    \end{equation}
    
for the RR fields. Explicitly we have
    \begin{align}
        \hat{F}_{2} &= F_{2} - B_{2}F_{0} \nonumber\\
        &=\frac{1}{4} e^{A(\rho)-\psi(\rho)}\sqrt{1-x(\rho)^{2}}\left[ \cos(\theta_{2})\Vol(\Sigma_{k}) -\Vol(S^{2}_{c}) \right] 
         +F_{0}\frac{e^{A(\rho)}}{4}\cos(\theta_{2})d\rho\wedge\left( d\phi_{2} -\frac{1}{k}\cos(\sqrt{k}\theta_{1})d\phi_{1} \right),\\
        \hat{F}_{4} &= F_{4}-F_{2}\wedge B_{2} + \frac{1}{2}B_{2}\wedge B_{2} F_{0}\nonumber\\
        &=\frac{1}{16}e^{2A(\rho)-\psi(\rho)} \sqrt{1-x(\rho)^{2}} d\rho\wedge d\phi_{2} \wedge \Vol(\Sigma_{k}).
    \end{align}

\subsubsection{Rewriting in terms of $\alpha(z)$}

Finally, it can be shown (as in \cite{Faedo:2019cvr}) that the 10D BPS equations can be solved in terms of just one function, $\alpha(z)$, provided $(A,x,\psi)$ are of the following form
    \begin{align}
        A(\rho)&= \frac{1}{2}\ln\,\left(8\pi \sqrt{2}\sqrt{-\frac{\alpha(z)}{\alpha''(z)}}\right),\\
        \psi(\rho)&= \frac{1}{4}\ln\left(e^{4\psi_{0}}
        \frac{\left(-\frac{\alpha(z)}{\alpha''(z)}\right)^{3}}{\Big( \alpha'(z)^{2}-2\alpha(z)\alpha'' (z)\Big)^{2}}\right),\\
        x(\rho)&= \sqrt{1+\frac{2\alpha(z)\alpha''(z)}{\alpha'(z)^2-2\alpha(z)\alpha''(z)}},
    \end{align}
    
where the coordinate $z$ is related to $\rho$ via the following change of coordinates
    \begin{equation}
        d\rho = \sqrt{\pi \sqrt{2}}\left(-\frac{\alpha''(z)}{\alpha(z)}\right)^{\frac{1}{4}}dz.
    \end{equation}
    
This is a solution of the 10D BPS equations provided $\alpha(z)$ satisfies
    \begin{equation}\label{eqF0AP}
        \alpha'''(z) = 2^{-\frac{1}{4}}\sqrt{\pi} e^{\psi_{0}}F_{0}.
    \end{equation}
    
In  terms of this new variable, the metric and background fields are the ones in \eqref{metric-z}-\eqref{NSRRforms}.

\section{Numerical solution of the BPS system}\label{appendixtwo}

Here we give a detailed derivation of the numerical solutions that describe the flow from AdS$_{7}$ to AdS$_{5}\times H^{2}$. The starting point is to consider a linear perturbation around the IR fixed point and use that as initial conditions for the numerical solution.

\subsection{Infrared Fixed Point}

We are looking for solutions to the BPS equations for $f(r)$, $h(r)$ and $\Phi(r)$, which we quote here for convenience,
    \begin{align}
        f' &= \frac{m}{2}\,e^{-2\Phi},\label{BPSf}\\
        h' &= \frac{1}{2}\left( \frac{1}{k}e^{-2h}+m\,e^{-2\Phi} \right),\label{BPSh}\\
        \Phi' &= \left(-1+\frac{1}{4k}e^{-2h} + m\,e^{-2\Phi}\right).
    \end{align}

that have constant $h(r)$ and $\Phi(r)$. This will correspond to the IR fixed point, and we note that it exists only for $k=-1$. The solutions 
    \begin{equation}
        f(r) = \frac{2}{3}r, \quad 
        h(r) = \frac{1}{2}\ln\left(\frac{3}{4}\right), \quad 
        \Phi(r) = \frac{1}{2}\ln\left(\frac{3m}{4}\right),
    \end{equation}

are exact solutions to the BPS equations, and the spacetime obtained from this corresponds to AdS$_{5}\times H^{2}$.

\subsection{Linear Perturbations}

Here we consider linear deviations from the IR fixed point
    \begin{align}
        f(r) &= \frac{2}{3}r + \epsilon ~a(r), \label{linear1}\\
        h(r) &= \frac{1}{2}\ln\left(\frac{3}{4}\right) + \epsilon~ b(r),\\
        \Phi(r) &= \frac{1}{2}\ln\left(\frac{3m}{4}\right) + \epsilon ~c(r).\label{linear3}
    \end{align}

Replacing in the BPS equations and keeping only first order terms  in $\epsilon$, we obtain linear equations for the linear perturbations
    \begin{align}
        a' &= -\frac{4}{3} c,\\
        b' &= \frac{4}{3}(b-c),\\
        c' &= \frac{1}{3}(2b-8c).
    \end{align}

This systems has as solution
    \begin{align}
    a(r) &= \frac{C_{1}}{3}(\sqrt{7}+2) e^{-\frac{2}{3}(\sqrt{7}+1)r} 
        -\frac{ C_{2}}{3}(\sqrt{7}-2) e^{\frac{2}{3}(\sqrt{7}-1)r},\\
    b(r) &= C_{1} e^{-\frac{2}{3}(\sqrt{7}+1)r} 
        + C_{2} e^{\frac{2}{3}(\sqrt{7}-1)r},\\
    c(r) &= \frac{C_{1}}{2}(\sqrt{7}+3) e^{-\frac{2}{3}(\sqrt{7}+1)r} 
        -\frac{C_{2}}{2}(\sqrt{7}-3)  e^{\frac{2}{3}(\sqrt{7}-1)r}.
    \end{align}

We are interested in solutions to vanish for $r\rightarrow-\infty$, which is the location of the fixed point, hence we set $C_{1}=0$ and also without loss of generality we can set $C_{2}=1$.

Finally, to obtain the numerical solutions we use the perturbations around the fixed point \eqref{linear1}-\eqref{linear3} as initial conditions at $r=-8$. Figure \ref{Functionspdf} shows these numerical solutions.

\section{Analysis of the 4d QFT}\label{sec-analysisqft}
Referring to the four dimensional ${\cal N}=1$ quiver in Figure \ref{quiver4d}, we assign R-charges and anomalous dimensions for each of the fields. For the adjoint scalars $\Phi_i$, the vector multiplets $W_i$, the bifundamentals between gauge nodes $(Q,\tilde{Q})$ and the fundamentals $(q,\bar{q})$ we  propose
\begin{eqnarray}
& &\text{The R-charges:}~~ R[\Phi_i]=1,\;\;\;R[W_i]=1, \;\;\; R[Q]=R[\tilde{Q}]=R[q]=R[\tilde{q}]=\frac{1}{2}.\label{Rcharges}\\
& & \text{The anomalous dimensions:}~~\gamma_{\Phi_i}=1,\;\;\; \gamma_{Q}=\gamma_{\tilde{Q}}=\gamma_{q}=\gamma_{\tilde{q}}=-\frac{1}{2}.\label{anomalous}
\end{eqnarray}
Notice that the dimension of any combination of fields  $\cal{O}$ satisfy 
\begin{equation}
\text{dim}{\cal O}=\frac{3}{2} R_{{\cal O}}.\nonumber
\end{equation}
As should occur at conformal points. In particular the following superpotential terms are present,
\begin{equation}
{\cal W}\sim \mu_i \Phi_i\Phi_i + q_i\bar{q}_jq_j\bar{q}_i+\text{all other combinations}.\label{superpot}
\end{equation}
The R-charge of each of these possible terms is $R[{\cal W}]=2$ and their dimension $\text{dim}[{\cal W}]=3$.
The mass term for all the adjoint scalars decouples them from the IR dynamics. They do not participate in the quantities computed below. The quartic term, on the other hand, generates interactions between all the different $P^2(g-1)$ horizontal lines of the quiver, by closing loops using the flavour groups. As one can start seeing, the behaviour of these quivers is somewhat reminiscent of those in \cite{Itsios:2017cew}.
Let us calculate the beta functions and the R-symmetry anomaly for each gauge group of the quiver in Figure \ref{quiver4d}.
\subsubsection{ Beta functions}
For any particular node we use the NSVZ beta function $\beta\sim 3 N_c- N_{fi}(1-\gamma_i)$, we find
\begin{equation}
\beta_i\sim 3 N_i - (N_{i+1}+ N_{i-1}+ F_i)(1-(-\frac{1}{2})) = \frac{3}{2}\Big[2 N_i- F_i- N_{i+1}- N_{i-1} \Big]=0.\label{beta4}
\end{equation}
As we start with a balanced quiver in the UV (the six dimensional quiver must cancel gauge anomalies), the beta function in four dimensions vanishes for each gauge group. It is nice to see how the six dimensional anomaly transmuted into the four dimensional beta function condition for conformality.
\subsubsection{ R-symmetry anomaly}
We calculate using the expression $\Delta \Theta= T(R_i) R(f_i)$. We use that $T(adj_i)=2 N_i$ and $T(fund)=1$. We find
\begin{equation}
\Delta \Theta_i= 2 N_i \times 1 + (F_i + N_{i+1} + N_{i-1})\times 2\times (-\frac{1}{2})= 2 N_i-F_i- N_{i+1}- N_{i-1}=0.
\end{equation}
The comments written below eq.(\ref{beta4}) also apply here. Both these calculations add to the proposal that the quiver in Figure \ref{quiver4d} describes the dynamics of our AdS$_5$ fixed point.
\\
Other interesting quantities are the central charges $a$ and $c$. 
\subsection{Central charges}
They are defined as,
\begin{equation}
a=\frac{3}{32\pi}\Big[3 \Tr R^3 - \Tr R \Big],\;\;\;\; c=\frac{1}{32\pi}\Big[9 \Tr R^3 - 5\Tr R \Big].\label{a-c}
\end{equation}
Calculating explicitly for the quiver in Figure \ref{quiver4d}, we find
\begin{eqnarray}
& & \Tr R= P^2 (g-1)\Big[\sum_{j=1}^{P-1} (N_j^2-1)\times 1 + N_j F_j \times (-\frac{1}{2})\times 2 + \sum_{j=1}^{P-2} N_j N_{j+1}  \times (-\frac{1}{2})\times 2 \Big]=\nonumber\\
& &  P^2 (g-1)\Big[\sum_{j=1}^{P-1} (N_j^2-1 -N_j F_j) - \sum_{j=1}^{P-2} N_j N_{j+1}  \Big].\nonumber\\
 & & \Tr R^3=  P^2 (g-1)\Big[\sum_{j=1}^{P-1} (N_j^2-1)\times 1^3 + N_j F_j \times (-\frac{1}{2})^3\times 2 + \sum_{j=1}^{P-2} N_j N_{j+1}  \times (-\frac{1}{2})^3\times 2 \Big]=\nonumber\\
& & =P^2 (g-1)\Big[\sum_{j=1}^{P-1} (N_j^2-1 -\frac{1}{4} N_j F_j) - \frac{1}{4} \sum_{j=1}^{P-2} N_j N_{j+1}  \Big].\nonumber
\end{eqnarray}
Using eqs.(\ref{a-c}) we have,
\begin{eqnarray}
& & a=\frac{3}{16\pi} P^2(g-1)\Big[ \sum_{j=1}^{P-1} (N_j^2-1)  +\frac{1}{8} N_j F_j + \frac{1}{8} \sum_{j=1}^{P-2} N_j N_{j+1} \Big],\label{acentral}\\
& & c= \frac{1}{8\pi} P^2(g-1)\Big[ \sum_{j=1}^{P-1} (N_j^2-1)  +\frac{11}{16} N_j F_j + \frac{11}{16} \sum_{j=1}^{P-2} N_j N_{j+1} \Big].\label{ccentral}
\end{eqnarray}
To gain some intuition, we discuss explicitly two illustrative examples.

\subsubsection{Example 1}
In this case we take
\begin{equation}
N_j= j N,\;\;\;\; F_j= N P \delta_{j, P-1}.\nonumber
\end{equation}
We find, for large values of $N$ and $P$ (the holographic limit!)
\begin{equation}
a= \frac{9}{128 \pi}P^5 N^2 (g-1)\left( 1 + O(\frac{1}{P})\right) ,\;\;\;\; c \sim a.\label{a-cexample1}
\end{equation}
The comparison between eqs.(\ref{a-cexample1}) and (\ref{chol4dex1}) suggests that in the QFT side there must be other fields that whilst not contributing to the beta function and R-anomaly, it makes a contribution to the $a$ and $c$ central charges. The contribution of these fields, that we denote as ${\cal X}$ should modify the result in eq.(\ref{a-cexample1}) (at leading order in $P$) according to,
\begin{equation}
a= \frac{9}{128\pi} N^2 P^5 (g-1)(1+ \delta_{{\cal X}}),~~~ 15 \delta_{{\cal X}}=1.\label{final1}
\end{equation}

\subsubsection{ Example 2}
In this case we take
\begin{equation}
N_j=  N,\;\;\;\; F_j= N ( \delta_{j,1}+\delta_{j, P-1}).\nonumber
\end{equation}
We find, in the holographic limit,
\begin{equation}
a= \frac{27}{128 \pi}P^3 N^2 (g-1)\left( 1+ O(\frac{1}{P}) \right),\;\;\;\; c \sim a.\label{a-cexample2}
\end{equation}
The comparison between eqs.(\ref{a-cexample2}) and (\ref{chol4dex2}) suggests that in the QFT side there must be other fields (not contributing to the beta function and R-anomaly), but adding a contribution to the $a$ and $c$ central charges. The contribution of these fields, that we denote as ${\cal X}$ should be of the form (at leading order in $P$)
\begin{equation}
a= \frac{27}{128\pi} N^2 P^3 (g-1)(1+ \delta_{{\cal X}}),~~~ 3 \delta_{{\cal X}}=1.\label{final2}
\end{equation}
This suggest that in our definition of the QFT quiver, new fields should enter adding these small contributions to the $a$ central charge.
Indeed, the presence of 'flip fields' that couple to irrelevant operators of the baryonic type is common in field theories of the class $S_{k}$. These fields are gauge singlets and decouple at low energies. They do not contribute to the beta functions or R-symmetry anomaly, but do change the result of $a,c$. It would be nice to see that their contribution can be $\delta_{{\cal X}}$ given in eqs.(\ref{final1})-(\ref{final2}).


\begin{thebibliography}{99}

\small	
	


\bibitem{Maldacena:1997re} 
  J.~M.~Maldacena,
  ``The Large N limit of superconformal field theories and supergravity,''
  Int.\ J.\ Theor.\ Phys.\  {\bf 38}, 1113 (1999)
  [Adv.\ Theor.\ Math.\ Phys.\  {\bf 2}, 231 (1998)]
  [hep-th/9711200].



\bibitem{oned}

Y.~Lozano, C.~Nunez, A.~Ramirez and S.~Speziali,
JHEP \textbf{03} (2021), 277
[arXiv:2011.00005 [hep-th]].

Y.~Lozano, C.~Nunez, A.~Ramirez and S.~Speziali,
JHEP \textbf{03} (2021), 145
[arXiv:2011.13932 [hep-th]].

Y.~Lozano, C.~Nunez and A.~Ramirez,
JHEP \textbf{04} (2021), 110
[arXiv:2101.04682 [hep-th]].





\bibitem{twod}
A.~Legramandi and N.~T.~Macpherson,
[arXiv:1912.10509 [hep-th]].

  C.~Couzens, H.~h.~Lam, K.~Mayer and S.~Vandoren,
  arXiv:1904.05361 [hep-th].


Y.~Lozano, N.~T.~Macpherson, C.~Nunez and A.~Ramirez,
JHEP \textbf{01}, 129 (2020)
[arXiv:1908.09851 [hep-th]].

 
\bibitem{Lozano:2019zvg}
Y.~Lozano, N.~T.~Macpherson, C.~Nunez and A.~Ramirez,
JHEP \textbf{01}, 140 (2020)
[arXiv:1909.10510 [hep-th]].


Y.~Lozano, N.~T.~Macpherson, C.~Nunez and A.~Ramirez,
Phys. Rev. D \textbf{101}, no.2, 026014 (2020)
[arXiv:1909.09636 [hep-th]].

Y.~Lozano, N.~T.~Macpherson, C.~Nunez and A.~Ramirez,
JHEP \textbf{12}, 013 (2019)
[arXiv:1909.11669 [hep-th]].
  %

C.~Couzens, Y.~Lozano, N.~Petri and S.~Vandoren,
Phys. Rev. D \textbf{105}, no.8, 086015 (2022)
[arXiv:2109.10413 [hep-th]].



\bibitem{threed}

  E.~D'Hoker, J.~Estes and M.~Gutperle,
  ``Exact half-BPS Type IIB interface solutions. II. Flux solutions and multi-Janus,''
  JHEP {\bf 0706}, 022 (2007)
  [arXiv:0705.0024 [hep-th]].
   E.~D'Hoker, J.~Estes, M.~Gutperle and D.~Krym,
  ``Exact Half-BPS Flux Solutions in M-theory. I: Local Solutions,''
  JHEP {\bf 0808}, 028 (2008)
  [arXiv:0806.0605 [hep-th]].


  B.~Assel, C.~Bachas, J.~Estes and J.~Gomis,
  ``Holographic Duals of D=3 N=4 Superconformal Field Theories,''
  JHEP {\bf 1108}, 087 (2011)
  [arXiv:1106.4253 [hep-th]].
  
  
  Y.~Lozano, N.~T.~Macpherson, J.~Montero and C.~Nunez,
  ``Three-dimensional $ \mathcal{N}=4 $ linear quivers and non-Abelian T-duals,''
  JHEP {\bf 1611}, 133 (2016)
  [arXiv:1609.09061 [hep-th]].
A.~Fatemiabhari and C.~Nunez,
[arXiv:2209.07536 [hep-th]].

P.~Merrikin and R.~Stuardo,
[arXiv:2112.10874 [hep-th]].


\bibitem{Coccia:2020wtk}
L.~Coccia and C.~F.~Uhlemann,
JHEP \textbf{06}, 038 (2021)
[arXiv:2011.10050 [hep-th]].


\bibitem{Akhond:2021ffz}
M.~Akhond, A.~Legramandi and C.~Nunez,
JHEP \textbf{11}, 205 (2021)
[arXiv:2109.06193 [hep-th]].









\bibitem{fourd}
  D.~Gaiotto and J.~Maldacena,
  ``The Gravity duals of N=2 superconformal field theories,''
  JHEP {\bf 1210}, 189 (2012)
  [arXiv:0904.4466 [hep-th]].

  R.~A.~Reid-Edwards and B.~Stefanski, jr.,
  ``On Type IIA geometries dual to N = 2 SCFTs,''
  Nucl.\ Phys.\ B {\bf 849}, 549 (2011)
  [arXiv:1011.0216 [hep-th]].
  
  O.~Aharony, L.~Berdichevsky and M.~Berkooz,
  ``4d N=2 superconformal linear quivers with type IIA duals,''
  JHEP {\bf 1208}, 131 (2012)
  [arXiv:1206.5916 [hep-th]].

Y.~Lozano and C.~Nunez,
JHEP \textbf{05}, 107 (2016)
[arXiv:1603.04440 [hep-th]].

  C.~Nunez, D.~Roychowdhury and D.~C.~Thompson,
  JHEP {\bf 1807}, 044 (2018)
  [arXiv:1804.08621 [hep-th]].
 
  \bibitem{4dss}
  C.~Nunez, D.~Roychowdhury, S.~Speziali and S.~Zacarias,
  Nucl.\ Phys.\ B {\bf 943}, 114617 (2019)
  [arXiv:1901.02888 [hep-th]].

























  
\bibitem{fived}
  E.~D'Hoker, M.~Gutperle, A.~Karch and C.~F.~Uhlemann,
  JHEP {\bf 1608}, 046 (2016)
  [arXiv:1606.01254 [hep-th]].
  

 E.~D'Hoker, M.~Gutperle and C.~F.~Uhlemann,
  Phys.\ Rev.\ Lett.\  {\bf 118}, no. 10, 101601 (2017)
  [arXiv:1611.09411 [hep-th]].
  E.~D'Hoker, M.~Gutperle and C.~F.~Uhlemann,
  ``Warped $AdS_6\times S^2$ in Type IIB supergravity II: Global solutions and five-brane webs,''
  JHEP {\bf 1705}, 131 (2017)
  [arXiv:1703.08186 [hep-th]].
  E.~D'Hoker, M.~Gutperle and C.~F.~Uhlemann,
``Warped $AdS_6\times S^2$ in Type IIB supergravity III: Global solutions with seven-branes,''
JHEP \textbf{11} (2017), 200
[arXiv:1706.00433 [hep-th]].
  
M.~Fluder and C.~F.~Uhlemann,
  ``Precision Test of AdS$_6$/CFT$_5$ in Type IIB String Theory,''
  Phys.\ Rev.\ Lett.\  {\bf 121}, no. 17, 171603 (2018)
  [arXiv:1806.08374 [hep-th]].

  O.~Bergman, D.~Rodriguez-Gomez and C.~F.~Uhlemann,
  JHEP {\bf 1808}, 127 (2018)
  [arXiv:1806.07898 [hep-th]].
 
    
\bibitem{Uhlemann:2019ypp}
  C.~F.~Uhlemann,
  ``Exact results for 5d SCFTs of long quiver type,''
  arXiv:1909.01369 [hep-th].

 
C.~F.~Uhlemann,
JHEP \textbf{09} (2020), 145
[arXiv:2006.01142 [hep-th]].


\bibitem{Legramandi:2021uds}
A.~Legramandi and C.~Nunez,
Nucl. Phys. B \textbf{974}, 115630 (2022)
[arXiv:2104.11240 [hep-th]].



\bibitem{sixd}
  S.~Cremonesi and A.~Tomasiello,
  JHEP {\bf 1605} (2016) 031
  [arXiv:1512.02225 [hep-th]].
  
F.~Apruzzi, M.~Fazzi, D.~Rosa and A.~Tomasiello,
[arXiv:1309.2949 [hep-th]].

  
  K.~Filippas, C.~Nunez and J.~Van Gorsel,
  JHEP {\bf 1906}, 069 (2019)
  [arXiv:1901.08598 [hep-th]].

O.~Bergman, M.~Fazzi, D.~Rodriguez-Gomez and A.~Tomasiello,
[arXiv:2002.04036 [hep-th]].

F.~Apruzzi, M.~Fazzi, A.~Passias, A.~Rota and A.~Tomasiello,
Phys.\ Rev.\ Lett.\  {\bf 115} (2015) no.6,  061601
[arXiv:1502.06616 [hep-th]].

\bibitem{Nunez:2018ags}
  C.~Nunez, J.~M.~Penin, D.~Roychowdhury and J.~Van Gorsel,
  JHEP {\bf 1806} (2018) 078
  [arXiv:1802.04269 [hep-th]].

\bibitem{Argyres:2022mnu}
P.~C.~Argyres, J.~J.~Heckman, K.~Intriligator and M.~Martone,
[arXiv:2202.07683 [hep-th]].


\bibitem{examples}

J.~M.~Maldacena and C.~Nunez,
Int. J. Mod. Phys. A \textbf{16}, 822-855 (2001)
[arXiv:hep-th/0007018 [hep-th]].
J.~M.~Maldacena and C.~Nunez,
Phys. Rev. Lett. \textbf{86}, 588-591 (2001)
[arXiv:hep-th/0008001 [hep-th]].
C.~Nunez, I.~Y.~Park, M.~Schvellinger and T.~A.~Tran,
JHEP \textbf{04}, 025 (2001)
[arXiv:hep-th/0103080 [hep-th]].
J.~P.~Gauntlett, D.~Martelli, J.~Sparks and D.~Waldram,
Class. Quant. Grav. \textbf{21}, 4335-4366 (2004)
[arXiv:hep-th/0402153 [hep-th]].
J.~P.~Gauntlett, D.~Martelli and D.~Waldram,
Phys. Rev. D \textbf{69}, 086002 (2004)
[arXiv:hep-th/0302158 [hep-th]].
F.~Benini and N.~Bobev,
JHEP \textbf{06}, 005 (2013)
[arXiv:1302.4451 [hep-th]].
I.~Bah, C.~Beem, N.~Bobev and B.~Wecht,
JHEP \textbf{06}, 005 (2012)
[arXiv:1203.0303 [hep-th]].
J.~M.~Maldacena and H.~S.~Nastase,
JHEP \textbf{09}, 024 (2001)
[arXiv:hep-th/0105049 [hep-th]].

\bibitem{Gaiotto:2009we}
D.~Gaiotto,
JHEP \textbf{08}, 034 (2012)
[arXiv:0904.2715 [hep-th]].

\bibitem{examples2}
D.~Gaiotto and S.~S.~Razamat,
JHEP \textbf{07}, 073 (2015)
[arXiv:1503.05159 [hep-th]].
S.~Franco, H.~Hayashi and A.~Uranga,
Phys. Rev. D \textbf{92}, no.4, 045004 (2015)
[arXiv:1504.05988 [hep-th]].
S.~S.~Razamat, E.~Sabag and G.~Zafrir,
JHEP \textbf{12}, 108 (2019)
[arXiv:1907.04870 [hep-th]].
I.~Bah, A.~Hanany, K.~Maruyoshi, S.~S.~Razamat, Y.~Tachikawa and G.~Zafrir,
JHEP \textbf{06}, 022 (2017)
[arXiv:1702.04740 [hep-th]].
S.~S.~Razamat, C.~Vafa and G.~Zafrir,
JHEP \textbf{04}, 064 (2017)
[arXiv:1610.09178 [hep-th]].




\bibitem{Bah:2021iaa}
I.~Bah, F.~Bonetti, E.~Leung and P.~Weck,
JHEP \textbf{09}, 197 (2022)
[arXiv:2112.07796 [hep-th]].
E.~Sabag and M.~Sacchi,
[arXiv:2208.03331 [hep-th]].
M.~Sacchi, O.~Sela and G.~Zafrir,
JHEP \textbf{05}, 053 (2022)
[arXiv:2111.12745 [hep-th]].


\bibitem{Razamat:2022gpm}
S.~S.~Razamat, E.~Sabag, O.~Sela and G.~Zafrir,
[arXiv:2203.06880 [hep-th]].




\bibitem{Baume:2021qho}
F.~Baume, M.~J.~Kang and C.~Lawrie,
[arXiv:2106.11990 [hep-th]].




\bibitem{Bah:2017wxp}
I.~Bah, A.~Passias and A.~Tomasiello,
JHEP \textbf{11}, 050 (2017)
[arXiv:1704.07389 [hep-th]].
 F.~Apruzzi, M.~Fazzi, A.~Passias and A.~Tomasiello,
[arXiv:1502.06620 [hep-th]].



\bibitem{Arkani-Hamed:2001wsh}
N.~Arkani-Hamed, A.~G.~Cohen, D.~B.~Kaplan, A.~Karch and L.~Motl,
JHEP \textbf{01}, 083 (2003)
doi:10.1088/1126-6708/2003/01/083
[arXiv:hep-th/0110146 [hep-th]].





\bibitem{Bobev:2017uzs}
N.~Bobev and P.~M.~Crichigno,
JHEP \textbf{12}, 065 (2017)
doi:10.1007/JHEP12(2017)065
[arXiv:1708.05052 [hep-th]].








 

 
 
 
   
\bibitem{Macpherson:2014eza} 
  N.~T.~Macpherson, C.~Nunez, L.~A.~Pando Zayas, V.~G.~J.~Rodgers and C.~A.~Whiting,
  JHEP {\bf 1502}, 040 (2015)
  [arXiv:1410.2650 [hep-th]].

\bibitem{Bea:2015fja} 
  Y.~Bea, J.~D.~Edelstein, G.~Itsios, K.~S.~Kooner, C.~Nunez, D.~Schofield and J.~A.~Sierra-Garcia,
  JHEP {\bf 1505}, 062 (2015)
  [arXiv:1503.07527 [hep-th]].
  














































\bibitem{GonzalezLezcano:2022mcd}
A.~Gonz\'alez Lezcano, J.~Hong, J.~T.~Liu, L.~A.~Pando Zayas and C.~F.~Uhlemann,
[arXiv:2207.09360 [hep-th]].

\bibitem{Legramandi:2021aqv}
A.~Legramandi and C.~Nunez,
JHEP \textbf{02}, 010 (2022)
[arXiv:2109.11554 [hep-th]].




\bibitem{Couzens:2022aki}
C.~Couzens, N.~T.~Macpherson and A.~Passias,
[arXiv:2209.15540 [hep-th]].

\bibitem{Passias:2015gya}
A.~Passias, A.~Rota and A.~Tomasiello,
JHEP \textbf{10} (2015), 187
[arXiv:1506.05462 [hep-th]].


\bibitem{Paredes:2004xw}
A.~Paredes,
[arXiv:hep-th/0407013 [hep-th]].

\bibitem{Faedo:2019cvr}
A.~F.~Faedo, C.~Nunez and C.~Rosen,
JHEP \textbf{03} (2020), 080
[arXiv:1912.13516 [hep-th]].

%
\bibitem{Itsios:2017cew}
G.~Itsios, Y.~Lozano, J.~Montero and C.~Nunez,
JHEP \textbf{09}, 038 (2017)
[arXiv:1705.09661 [hep-th]].
I.~Bah and N.~Bobev,
JHEP \textbf{08}, 121 (2014)
[arXiv:1307.7104 [hep-th]].
 
 









\end{thebibliography}
\end{document}